\pdfoutput=1
\documentclass[final,times,1p]{elsarticle}

\usepackage{graphics}
\usepackage{graphicx,stfloats}
\usepackage{color}

\usepackage{amssymb}
\usepackage{amsmath}
\usepackage{amsmath}
\usepackage{amsfonts}
\usepackage{amssymb}

\usepackage{amsthm}
\usepackage{amsmath}
\usepackage{soul}
\usepackage{xcolor}
\usepackage{commath}
\usepackage{latexsym}
\usepackage{subfig}
\usepackage{float}
\usepackage{url}
\usepackage{setspace}
\usepackage{multirow}
\usepackage{threeparttable}
\usepackage{booktabs}
\usepackage{textgreek}

\newcommand{\ode}[2]{\dfrac{\text{d} {#1}}{\text{d} {#2}}}  

\biboptions{sort&compress}

\journal{Combustion and Flame}

\begin{document}
	
\begin{frontmatter}
	
\title{Role of Instability on the Limits of Laterally Strained Detonation Waves}

\author[add1]{Qiang Xiao\corref{cor1}}
\ead{qxiao067@uottawa.ca}
\author[add1]{Matei I. Radulescu}

\address[add1]{Department of Mechanical Engineering, University of Ottawa, Ottawa, \\Ontario K1N 6N5, Canada}
\cortext[cor1]{Corresponding author}

\begin{abstract}
	
The present work examines the role of instability and diffusive phenomena in controlling the limits of detonations subject to \textcolor{black}{lateral strain}. Experiments were conducted in mixtures with varying levels of cellular instability, i.e., stoichiometric methane-oxygen (CH$_4$/2O$_2$), ethylene-oxygen (C$_2$H$_4$/3O$_2$), and ethane-oxygen (C$_2$H$_6$/3.5O$_2$). These detonations were propagated in channels with exponentially enlarging cross-sections, following the recent works of Radulescu \& Borzou (2018) and Xiao \& Radulescu (2020).  Steady detonation waves were obtained at the macro-scale, with the near-limit reaction zone structures characterized by significant unreacted gas pockets. The turbulent flame burning velocity of these pockets was evaluated to be 30 m/s to 70 m/s, which is larger than the theoretical laminar value by a factor of 2 to 7 and smaller than the CJ deflagration speed by a factor of 2 to 3. For all the mixtures tested, the characteristic $D-K$ relationships, relating the detonation mean propagation speed with lateral flow divergence, were obtained directly from experiments and as well from the generalized ZND model with lateral strain rates using \textcolor{black}{detailed chemical kinetics}. The results showed that the degree of \textcolor{black}{departure} between experiments and the theoretical predictions increases significantly with the detonation instability level. As compared to the laminar ZND wave, the more unstable detonations are much more detonable than the more stable detonations, with substantially larger limiting divergence rates and maximum velocity deficits. Such enhanced detonability with detonation instability can be manifested in the significantly enhanced global rates of energy release with the notably suppressed thermal character of ignition for the more unstable detonations. This globally enhanced burning mechanism is found to be realized by the intensified auto-ignition assisted by the turbulent diffusive burning of the unreacted gas pockets, substantially shortening the characteristic reaction zone lengths. Finally, the relatively good universality of the $D/D_{CJ}-K_{\text{eff}} \lambda$ relations implicitly suggests that cell size is also a function of the detonation wave stability, since it is controlled by the reaction zone thickness of unstable detonations.
\end{abstract}

\begin{keyword}
Detonation instability  \sep Unreacted gas pocket  \sep Diffusive phenomenon \sep  \textcolor{black}{Lateral strain} \sep Detonation limits
\end{keyword}

\end{frontmatter}

\section{Introduction}
\label{Introduction}

Very recently, Xiao and Radulescu \cite{xiao2019} performed experiments with hydrogen-oxygen-argon (H$_2$/O$_2$/Ar) cellular detonations in exponentially diverging channels. They found that, despite the cellular structures, the detonation dynamics can be excellently predicted from the generalized Zeldovich-von Neumann-Doering (ZND) model with lateral strain rates \cite{klein1995} using \textcolor{black}{detailed chemical kinetics}. However, this excellent agreement departs from the earlier experiments of Radulescu and Borzou \cite{radulescu2018} in \textcolor{black}{more unstable} detonations, where the experiments disagreed with the theoretical predictions. The question thus arises whether there is the link of \textcolor{black}{departure} between experiments and the steady 1D ZND model predictions to detonation instability. The present work addresses this question.

In practice, detonations usually propagate in the presence of \textcolor{black}{lateral strain} \cite{fickett1979detonation, lee_2008}. For example, in either weakly confined (e.g., see Refs.\ \cite{dabora1965, reynaud2017a, houim2017, cho2017, mi2018b, chiquete2019}) or varying-cross-section (e.g., see Refs.\ \cite{xiao2019, radulescu2018,nagura2013}) or curved geometries (e.g., see Refs.\ \cite{kudo2011oblique,nakayama2012,nakayama2013, rodriguez2019, short2019a}), detonations tend to be curved with the flow experiencing lateral divergence in the reaction zone. A typical case can be seen in rotating detonation engines (RDEs) \cite{schwer2011numerical, lu2014, paxson2018examination, anand2019rotating}, where detonations are weakly confined by the burned products of the previous rotation cycle and the flow behind \textcolor{black}{the front} locally experiences side expansions, resulting in \textcolor{black}{lateral strain}. The other important scenario is related with detonations in small-sized confined geometries, such as narrow channels or tubes, where detonations are subject to significant losses induced by boundary layers \cite{zeldovich1950, wood1954a, fay1959, tsuge1971, dove1974, murray1985,gelfand1991, agafonov1994, dionne2000, ishii2002, radulescu2002failure, radulescu2003propagation,chao2009, kitano2009, camargo2010, ishii2011a, zhang2015, gao2016}. In the shock-attached frame of reference, the boundary layer acts as a mass sink from the inviscid core flow \cite{fay1959} and results in the flow diverging in reaction zones \cite{chinnayya2013, sow2019}, thereby giving rise to \textcolor{black}{lateral strain}. This \textcolor{black}{boundary-layer-induced} lateral flow divergence can thus be \textcolor{black}{modelled} by the negative displacement of boundary layers \cite{fay1959}. \textcolor{black}{Lateral strain has} been generally known to significantly impact the detonation dynamics, i.e., decreasing the propagation speeds lower than the theoretical Chapman-Jouguet (CJ) \textcolor{black}{velocities}  \textcolor{black}{\cite{fay1959, dove1974, murray1985, radulescu2002failure, chao2009, kitano2009, camargo2010, ishii2011a, zhang2015, gao2016, xiao2019effect}}, increasing the limit pressures \cite{dove1974, radulescu2002failure, chao2009, camargo2010} as well as cell sizes \cite{strehlow1967a, ishii2011a, monwar2007a, chiquete2019, reynaud2017a, xiao2019effect}. 

Extensive efforts have been made in quantifying the response of detonation dynamics to \textcolor{black}{lateral strain} for detonations in narrow channels or small tubes (e.g., see Refs. \textcolor{black}{\cite{fay1959, dove1974, murray1985, radulescu2002failure, chao2009, kitano2009, camargo2010, ishii2011a, zhang2015, gao2016}}). The experimentally obtained detonation mean propagation speed dependence on ``lateral flow divergence'' was then compared to the extended ZND models \cite{fay1959, dove1974, murray1985, klein1995}. Note that in these works, the ``lateral flow divergence'' was interpreted in terms of the characteristic geometry length scale (e.g., the channel depth or the tube diameter) normalized by the length scale of the kinetics, such as the detonation cell size \cite{strehlow1967a} or the ZND detonation induction zone length \cite{radulescu2002failure, klein1995}. In general, all the comparisons showed that the steady 1D ZND model can relatively well predict the dynamics of weakly unstable detonations with regular cellular structures, while much worse for the highly unstable detonations with more irregular cellular structures. Specifically in terms of detonation limits, as compared to the steady ZND counterpart, unstable detonations in experiments can propagate at substantially larger limiting flow divergence with much larger velocity deficits \cite{radulescu2002failure, radulescu2003propagation, mazaheri2015experimental}. Despite these findings, whether there is a correlation between the \textcolor{black}{departure} and detonation instability is still not clear. Moreover, no effective strategies have been proposed in improving the theoretical predictions of the unstable detonation dynamics. 

In an attempt to \textcolor{black}{tackle} these questions, Radulescu and Borzou \cite{radulescu2018} recently formulated a novel experimental technique involving a pair of exponentially diverging ramps. The key feature of the ramp is keeping the logarithmic area divergence rate  as a constant, which can enable the macro-scale detonations to propagate in quasi-steady conditions with a constant mean lateral strain rate \cite{radulescu2018, xiao2019}. As a result, the unique characteristic relationships between the detonation mean propagation speed and its lateral flow divergence can be precisely obtained in an unambiguous manner, thus permitting meaningful comparisons between experiments and the steady ZND model predictions. In their works, two mixtures of different detonation instability were tested, i.e., the weakly unstable argon-diluted acetylene-oxygen (2C$_2$H$_2$/5O$_2$/21Ar) and highly unstable propane-oxygen (C$_3$H$_8$/5O$_2$) detonations. Again, they found \textcolor{black}{a} much poorer predictability of the unstable C$_3$H$_8$/5O$_2$ detonation dynamics than that of the more stable 2C$_2$H$_2$/5O$_2$/21Ar detonations. Besides these comparisons, they extracted the empirical reaction models from the experimental data for the global average description of the detonation dynamics. The potential usefulness of these calibrated global rate laws has been demonstrated by Radulescu \cite{radulescu2017icders} in predicting the initiation and failure phenomena in relevant C$_3$H$_8$/5O$_2$ detonation experiments \cite{zhang2013a}. On the other hand, \textcolor{black}{for highly unstable  C$_3$H$_8$/5O$_2$ detonations, the calibrated activation energy was found to be significantly lower and the reaction kinetics faster in comparison to the ZND model with the underlying chemistry. This fact highlights} the role of diffusive processes, involved in the burn-out of unreacted gas pockets, in suppressing the thermal character of the ignition mechanism \cite{radulescu2007, maxwell2017}. Such diffusive phenomenon is essential in resolving the ``detonation paradox'' that inviscid detonation simulations predict the incorrect trend for the role of instability in real gaseous cellular detonations \cite{radulescu2018c}. 

Therefore, by further extending the well-posed exponential ramp technique, the present study aims to examine how the instability and diffusive processes affect detonation limits subject to \textcolor{black}{lateral strain} in a range of mixtures with varying levels of cellular instability, i.e., stoichiometric methane-oxygen (CH$_4$/2O$_2$), ethylene-oxygen (C$_2$H$_4$/3O$_2$), and ethane-oxygen (C$_2$H$_6$/3.5O$_2$). Particularly, it is of interest to establish whether there is the link of \textcolor{black}{departure} between \textcolor{black}{experiments} and the generalized ZND model predictions to detonation instability. Finally, useful empirical global reaction models are presented for capturing the detonation dynamics.

\section{Experimental Details}
\label{Experimental Details}

\begin{figure}[]
	\centering
	{\includegraphics[width=1.0\columnwidth]{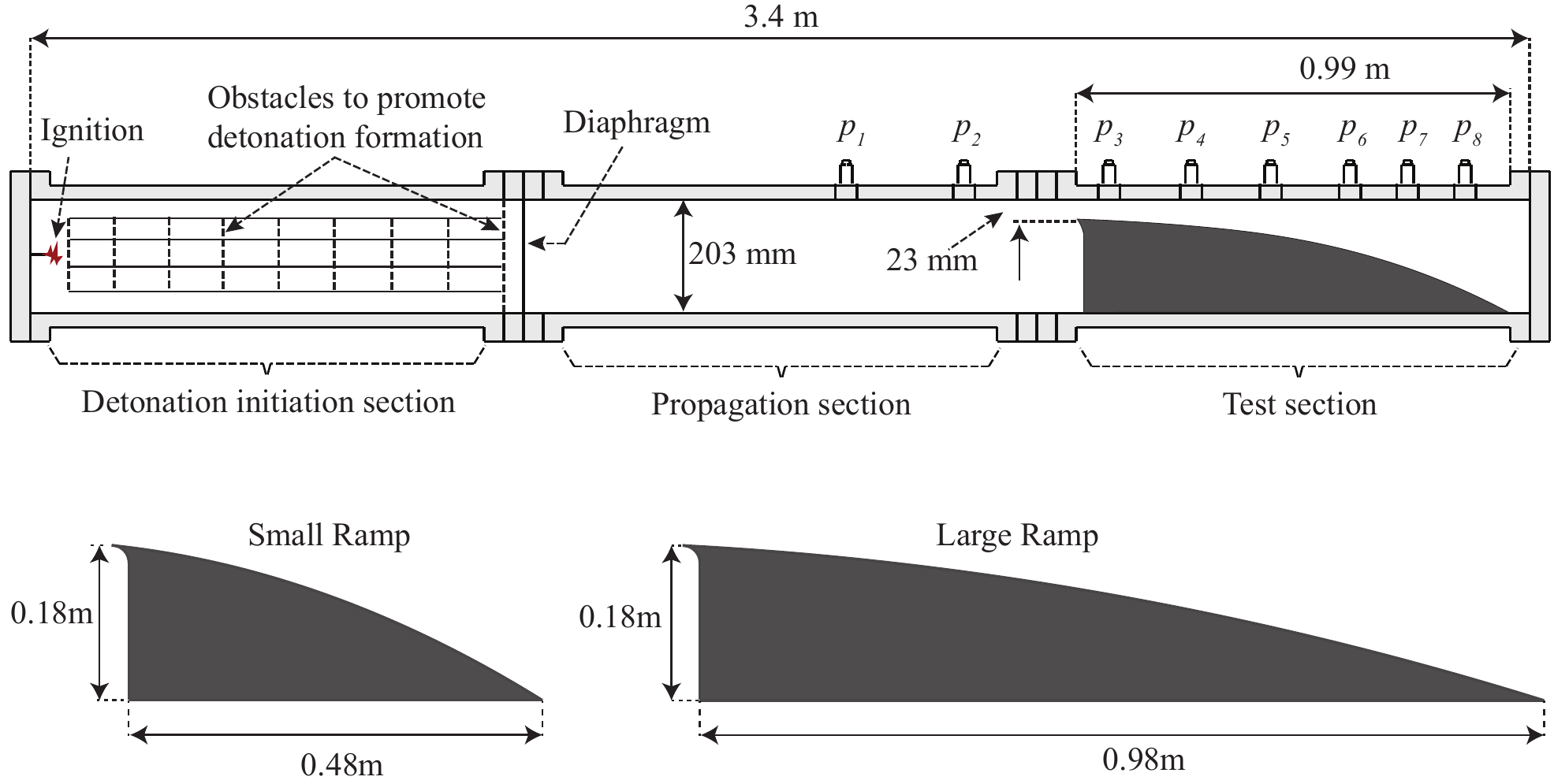}}
	\caption{Experimental set-up for diverging detonation experiments. } \label{Experimental-setup}  
\end{figure}

Experiments were conducted in a 3.4 m long aluminium rectangular channel with an internal height and width of 203 mm and 19 mm, respectively. A sketch of the experimental set-up is shown in Fig.\ \ref{Experimental-setup}, which is the same as that adopted by Xiao and Radulescu \cite{xiao2019} and Radulescu and Borzou \cite{radulescu2018}. The shock tube consists of three parts, a detonation initiation section, a propagation section, and a test section. In the initiation part, the mixture was ignited by a high voltage igniter (HVI), which could store up to 1000 J with the deposition time of 2 $\mu$s. Mesh wires were inserted  for facilitating the detonation formation. Three 113B24 and five 113B27 piezoelectric PCB pressure sensors ($p_1 \sim p_8$), as shown in Fig.\ \ref{Experimental-setup}, were mounted flush on the top wall of the shock tube for recording pressure signals. 

In the test section, two different polyvinyl-chloride (PVC) ramps were used in our experiments. The key characteristic of the ramp is the property of retaining a constant logarithmic area divergence rate ($K=\frac{d(lnA(x))}{dx}$) of the cross-sectional area $A(x)$. This important feature enables detonations at the macro-scale to propagate in quasi-steady conditions \cite{radulescu2018, xiao2019}. For the large ramp,  the logarithmic area divergence rate \textcolor{black}{was} 2.17 $\textnormal{m}^{-1}$, while for the small \textcolor{black}{one, the} rate was 4.34 $\textnormal{m}^{-1}$. At the entrance, a protruded rounded tip was kept for minimizing the effects of shock reflection on the detonation front. The initial gap between the ramp tip and the top wall of the channel \textcolor{black}{was} 23 mm in height. The height between the exponentially curved wall and the top wall is given by $y_{\textrm{{wall}}} = y_{0} e^{Kx}$ for $x>0$ and by $y_{\textrm{{wall}}} = y_{0}$ = 23 mm otherwise. 

The test mixtures  were stoichiometric methane-oxygen (CH${_4}$/2O${_2}$), ethylene-oxygen (C${_2}$H${_4}$/3O${_2}$), and ethane-oxygen (C${_2}$H${_6}$/3.5O${_2}$).  Each of them was prepared in a separate mixing tank by the method of partial pressures and was then left to mix for more than 24 hours. Before filling with the test mixture in every single experiment, the shock tube was evacuated  below the absolute pressure of 90 Pa. The mixture was then introduced into the shock tube through both ends of the tube at the desired initial pressure with an accuracy of 70 Pa. Both the schlieren \cite{bhattacharjee2013} and large-scale shadowgraph \cite{dennis2014} techniques were alternatively adopted for visualizing the detonation evolution process along the exponential ramp. The Z-type schlieren was set up with a vertical knife edge using a light source of 360 W. The resolution of the high-speed camera was 384$\times$288 $\textnormal{px}^{2}$ with the framing rate of 77481 fps (about 12.9 $\mu $s for each interval). For the shadowgraph photos, the resolution of the high-speed camera was 1152$\times$256 $\textnormal{px}^{2}$ with the framing rate of 42049 fps (about 23.78 $\mu $s for each interval). The light source for the shadowgraph system was provided by a 1600 W Xenon arc \textcolor{black}{lamp}. The exposure time for the schlieren and shadowgraph setup was 0.44 $\mu $s and 1.0 $\mu $s, respectively.

\section{Experimental Results}
\subsection{Detonation propagation characteristics}
\label{Detonation propagation characteristics}
\subsubsection{Methane-oxygen detonations}

\begin{figure}[]
	\centering
	{\includegraphics[width=1.0\columnwidth]{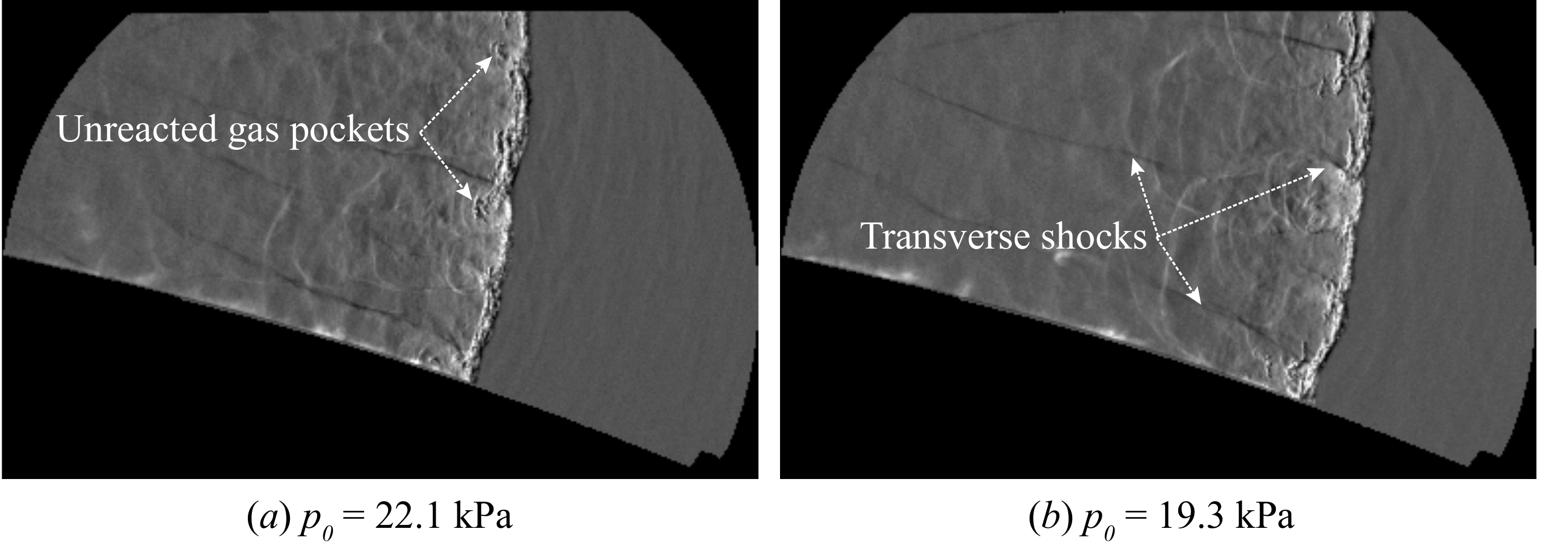}}
	\caption{Structures of CH$_4$/2O$_2$ detonations propagating along the large ramp at relatively high initial pressures.} \label{CH4-2O2-Structure1}  
\end{figure}

\begin{figure}[]
	\centering
	{\includegraphics[width=0.95\columnwidth]{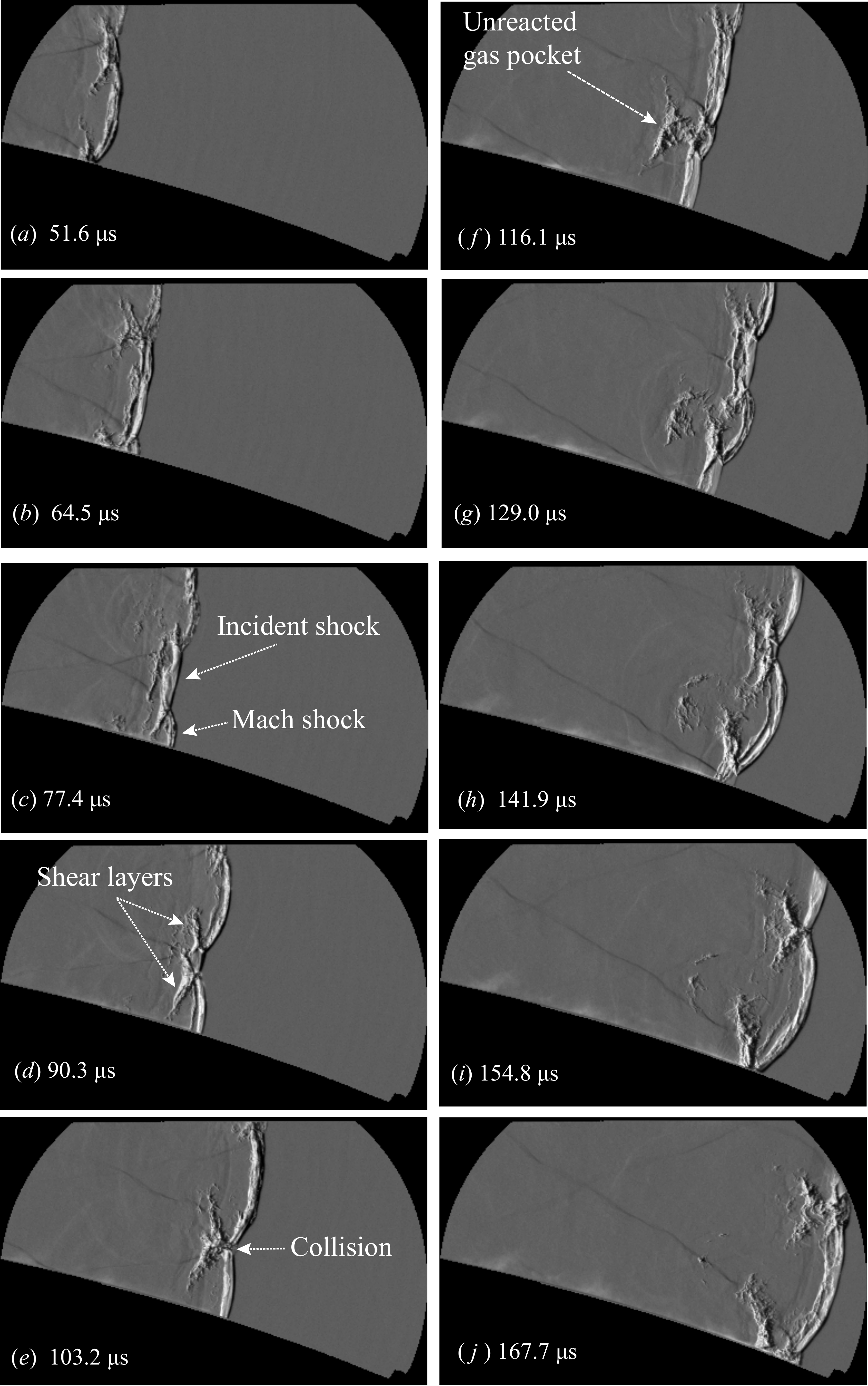}}
	\caption{Evolution of CH${_4}$/2O$_2$ detonations along the large ramp at the initial pressure of 12.9 kPa.  } \label{CH4-2O2-Evolution-1}  
\end{figure}

Figure\ \ref{CH4-2O2-Structure1} shows the reaction zone structures of methane-oxygen (CH${_4}$/2O$_2$) detonations along the large ramp at initial pressures relatively well above the propagation limit of $p_c \approx 12.0$ kPa, below which detonations fail to self-sustain. The detonation propagated from left towards right.  As a result of the geometrical area divergence, detonation fronts are curved. The other main features are the long-tailed transverse shocks, the fine-scale dense unreacted gas pockets breaking away from the very rough front, and also the trailing scattered pressure pulses behind the spotty frontal structure.  These characteristics have been previously reported by Radulescu et al. \cite{radulescu2005, radulescu2007} and  Kiyanda and Higgins \cite{kiyanda2013} for methane-oxygen detonations in straight channels.  One can have a clearer observation of the formation process of the unreacted gas pockets or islands in the propagation of much lower initial pressure detonations, as shown in Fig.\ \ref{CH4-2O2-Evolution-1}. It illustrates the detailed \textcolor{black}{interactions} of triple points in the three-mode (i.e., one cell and half) detonation evolution process along the large ramp. Due to the longer ignition delays for the shocked gases processed by the weaker incident shock, the observed reaction zones are thicker than that behind the stronger Mach shock. The tongue-like region of unreacted gases accumulated along the shear layers increases until the collision of triple points, giving rise to the dense unburned gas island peeled away from the main front, as can be seen from Fig.\ \ref{CH4-2O2-Evolution-1}$c$ to Fig.\ \ref{CH4-2O2-Evolution-1}$f$. In the experiment, this unreacted gas pocket was consumed within five frames, i.e., approximately 65 $\mu$s, from  Fig.\ \ref{CH4-2O2-Evolution-1}$e$ to Fig.\ \ref{CH4-2O2-Evolution-1}$j$. Noteworthy is that this burn-out time of unreacted pockets is approximately the timescale of the detonation cell, consistent with the previous observations \cite{radulescu2005, radulescu2007, kiyanda2013, maxwell2017}.
       
\begin{figure}[]
	\centering
	{\includegraphics[width=0.95\textwidth]{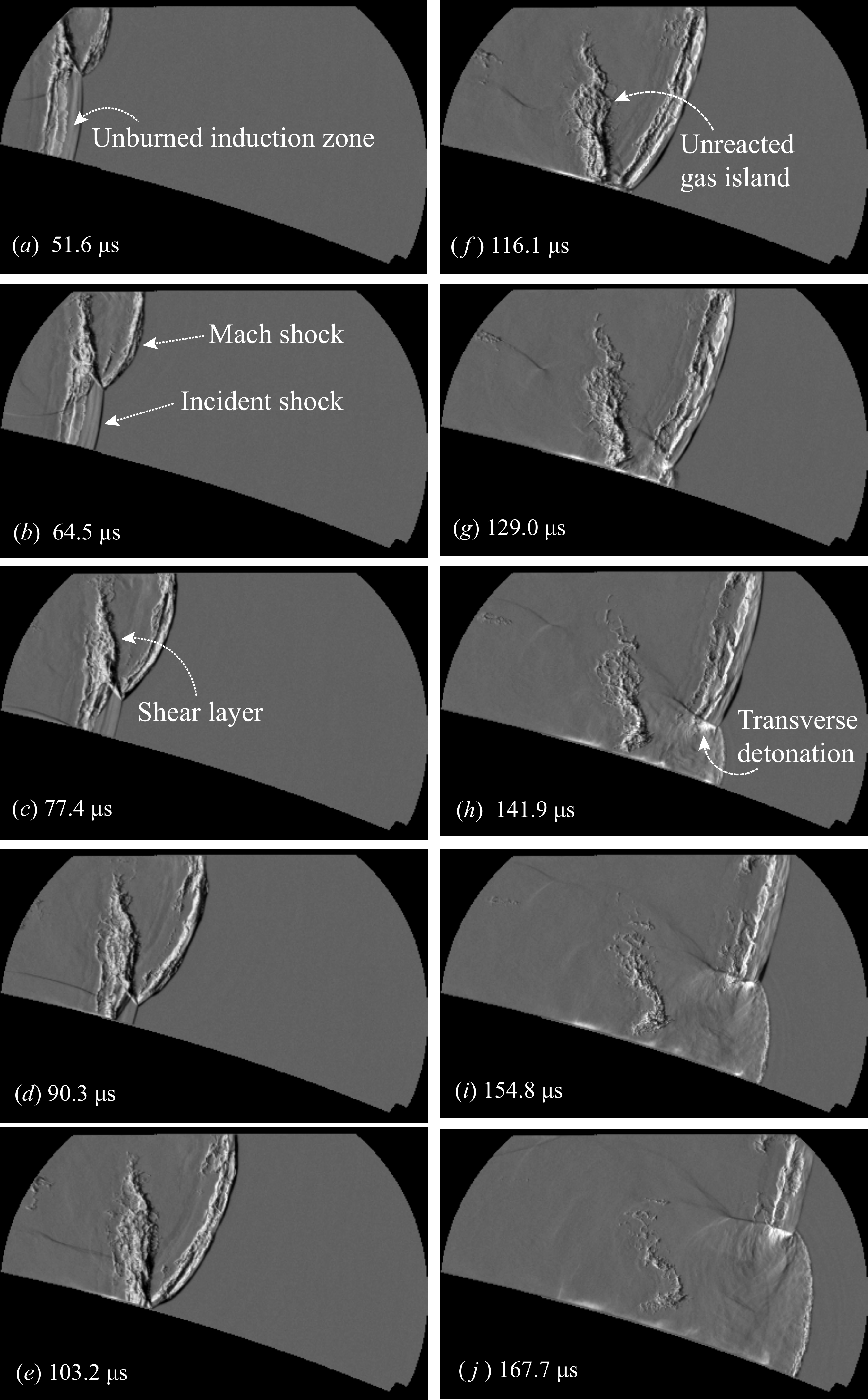}}
	\caption{Evolution of CH${_4}$/2O$_2$ detonations along the large ramp at the initial pressure of 12.7 kPa.  } \label{CH4-2O2-Evolution-2}  
\end{figure}

Such behavior can be further confirmed in the single-mode (or single-head, i.e., half cell) detonation propagation, as shown at some length in Fig.\ \ref{CH4-2O2-Evolution-2}. The prominent feature of the detonation in Fig.\ \ref{CH4-2O2-Evolution-2}$a$ is the highly turbulent tongue-shaped structure (along the shear layer) engulfing the unreacted gases in the large induction zone behind the weaker incident shock. This structure is qualitatively similar to that observed by Kiyanda and Higgins \cite{kiyanda2013}. As the triple point propagated downward from Fig.\ \ref{CH4-2O2-Evolution-2}$a$ to Fig.\ \ref{CH4-2O2-Evolution-2}$e$, the tongue of dense unburned gases along the turbulent shear layer was noticeably enlarged, and finally pinched off from the main front as a separate island during the collision of the triple point with the bottom wall, as can be seen in Fig.\ \ref{CH4-2O2-Evolution-2}$f$. In the meantime, a transverse detonation of intense luminosity, as documented numerically by Gamezo et al. \cite{gamezo2000} and experimentally by Bhattacharjee et al. \cite{bhattacharjee2013},  was generated and then rapidly \textcolor{black}{burned} the unreacted gases behind the upper incident shock (see frames from Fig.\ \ref{CH4-2O2-Evolution-2}$f$ to Fig.\ \ref{CH4-2O2-Evolution-2}$j$). Such formation mechanism of the transverse detonation in a relatively large unburned induction zone through the reflection of the transverse wave from the wall is similar to that observed \textcolor{black}{recently} by Xiao and Radulescu \cite{xiao2019} in the mixture of 2H$_2$/O$_2$/7Ar. Starting in Fig.\ \ref{CH4-2O2-Evolution-2}$f$, the unreacted gas island was observed to be fully consumed within eight frames, i.e., 103 $\mu$s, which is the timescale of the detonation cell. Nevertheless, the propagation speeds of incident shocks (close to the bottom curved wall) from Fig.\ \ref{CH4-2O2-Evolution-2}$a$ to Fig.\ \ref{CH4-2O2-Evolution-2}$e$ were measured decreasing from  0.55$D_{CJ}$ to 0.45$D_{CJ}$, whose average is 0.5$D_{CJ}$. Correspondingly, its post-shock ignition delay (at the average speed of 0.5$D_{CJ}$) was calculated to be 10.3 seconds, which is five orders of magnitude longer than the experimentally estimated burn-out time of the unreacted gases processed by this \textcolor{black}{shock.} Note that the theoretical ignition time was evaluated using the typical \textcolor{black}{constant-volume} explosion method with the detailed San Diego reaction mechanism \cite{Williams2014}. Therefore, it is clear that the shock-induced auto-ignition theory alone cannot account for the much faster reaction mechanism of the unreacted fuel pockets observed in these highly unstable detonation experiments, as already noted in previous works \cite{radulescu2005, radulescu2007, kiyanda2013, maxwell2017}.     

\begin{figure}[]
	\centering
	{\includegraphics[width=1.0\textwidth]{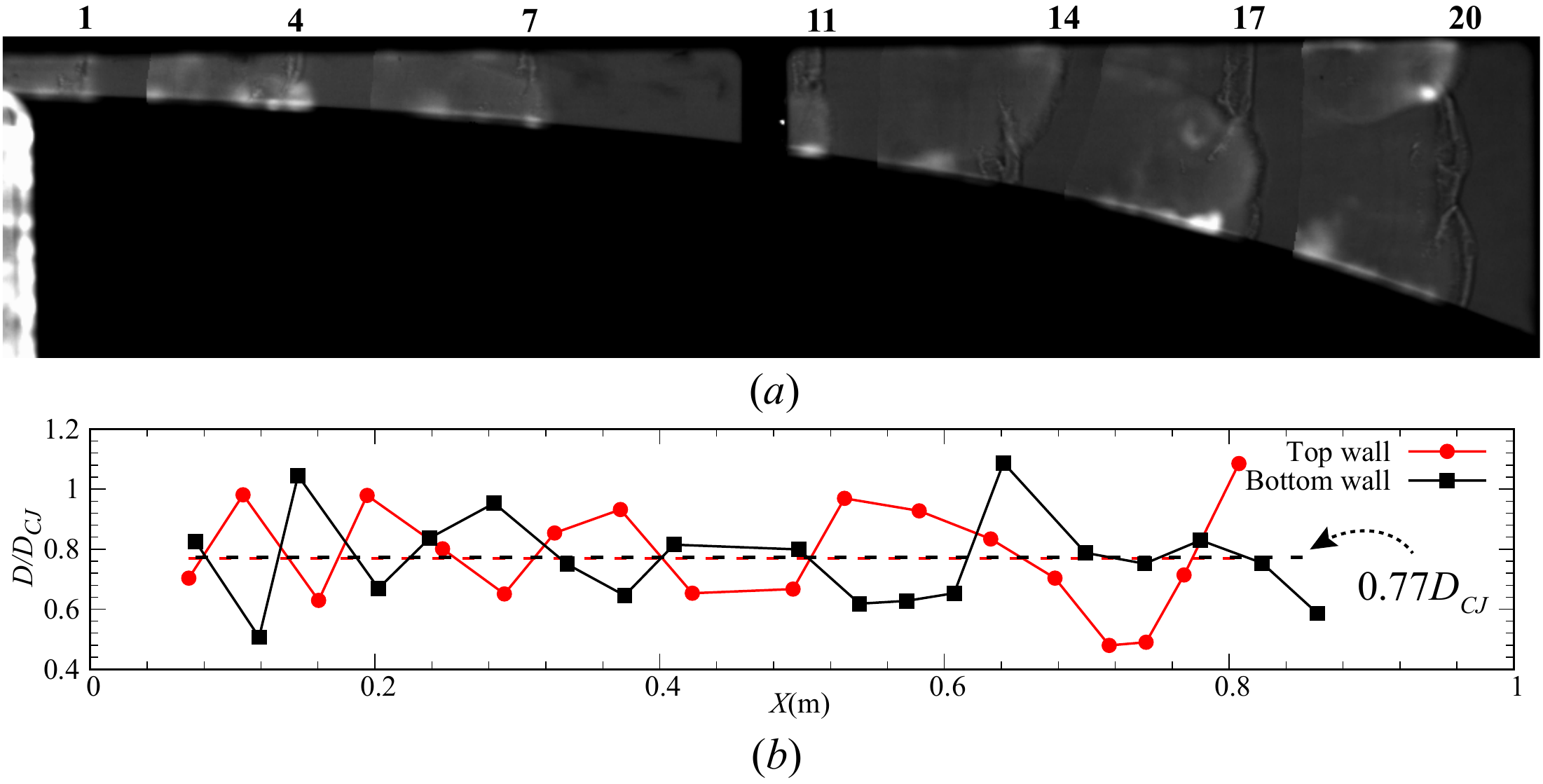}}
	\caption{\textcolor{black}{Superposition of detonation fronts $\left(a\right)$ and the corresponding local speed profiles $\left(b\right)$ at different time instants along the large ramp in the mixture of CH${_4}$/2O$_2$ at an initial pressure of $p_0 = 13.5$ kPa. Numbers in $\left(a\right)$ represent the frame sequence. The dashed red and black lines in $\left(b\right)$ denote the mean propagation speeds over the whole ramp for the top and bottom walls, respectively. Video animations were provided as Supplemental material illustrating the detailed evolution process. }  } \label{CH4-limit}  
\end{figure}

\textcolor{black}{In addition, we also obtained the local speed profiles of the near-limit detonation over the whole ramp, as shown in Fig.\ \ref{CH4-limit}. The locally averaged speeds both along the top and bottom walls (in Fig.\ \ref{CH4-limit}$b$) were calculated from the corresponding shadowgraph photos (partly superimposed in Fig.\ \ref{CH4-limit}$a$), with the distance between every two neighboring frames divided by their time interval $\Delta t$. It is clear that the local speed profile in Fig.\ \ref{CH4-limit}$b$ exhibits periodic fluctuations typical of cellular dynamics, which ranges from the maximum of 1.2 $D_{CJ}$ to the minimum of 0.5 $D_{CJ}$. Despite these significant local variations, detonations in each cellular cycle appeared to propagate at approximately the same mean propagation speed of 0.77$D_{CJ}$ along both the top and bottom curved walls. This fact has been detailed recently by Xiao and Radulescu \cite{xiao2019} and Radulescu and Borzou \cite{radulescu2018}. Therefore, it is still meaningful to assume a constant mean propagation speed for the near-limit detonations inside the exponentially diverging channels.}

\subsubsection{Ethylene-oxygen and ethane-oxygen detonations}

\begin{figure}[]
	\centering
	{\includegraphics[width=1.0\textwidth]{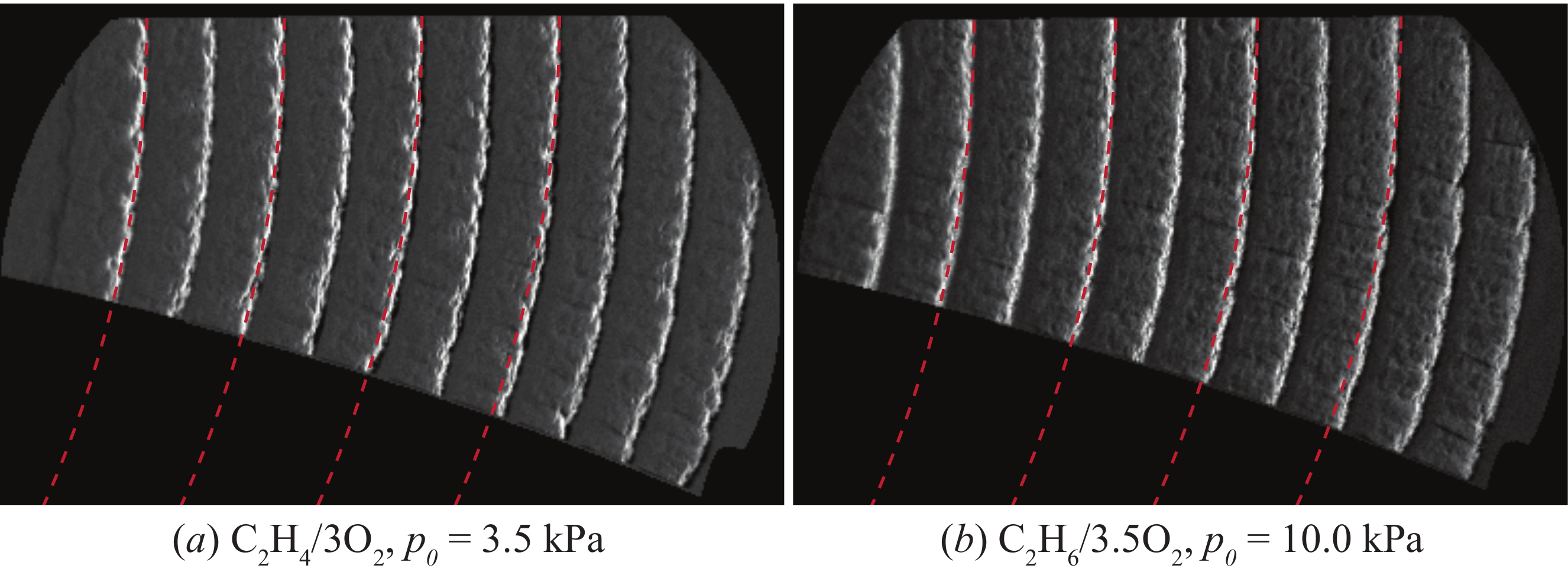}}
	\caption{Superposition of detonation fronts near the end of the large ramp in mixtures of C${_2}$H${_4}$/3O$_2$ and C${_2}$H${_6}$/3.5O$_2$, respectively, at initial pressures well above the limit. The red lines denote arcs of circles with the expected curvature from the quasi-1D approximation. } \label{Superposition1}  
\end{figure}

Figure\ \ref{Superposition1} shows the superimposed schlieren photos of detonation fronts near the end of the large ramp at different instants in mixtures of ethylene-oxygen (C${_2}$H${_4}$/3O$_2$) and ethane-oxygen (C${_2}$H${_6}$/3.5O$_2$), respectively. They are at initial pressures well above the limit of $p_c \approx 1.0$ kPa for C${_2}$H${_4}$/3O$_2$ and $p_c \approx 4.0$ kPa for C${_2}$H${_6}$/3.5O$_2$.  The detonation propagated from left towards right. The detonation front acquired a large number of very small-sized cellular structures, and was noticeably curved with a characteristic curvature due to the cross-sectional area divergence. In general, both the  C${_2}$H${_4}$/3O$_2$ and C${_2}$H${_6}$/3.5O$_2$ detonation fronts are qualitatively similar, while appear to be less turbulent than the very ``rough" ones of CH${_4}$/2O$_2$ in Fig.\ \ref{CH4-2O2-Structure1}. The C${_2}$H${_4}$/3O$_2$ and C${_2}$H${_6}$/3.5O$_2$ detonations are organized with much thinner characteristic reaction zones.

\begin{figure}[]
	\centering
	{\includegraphics[width=1.0\textwidth]{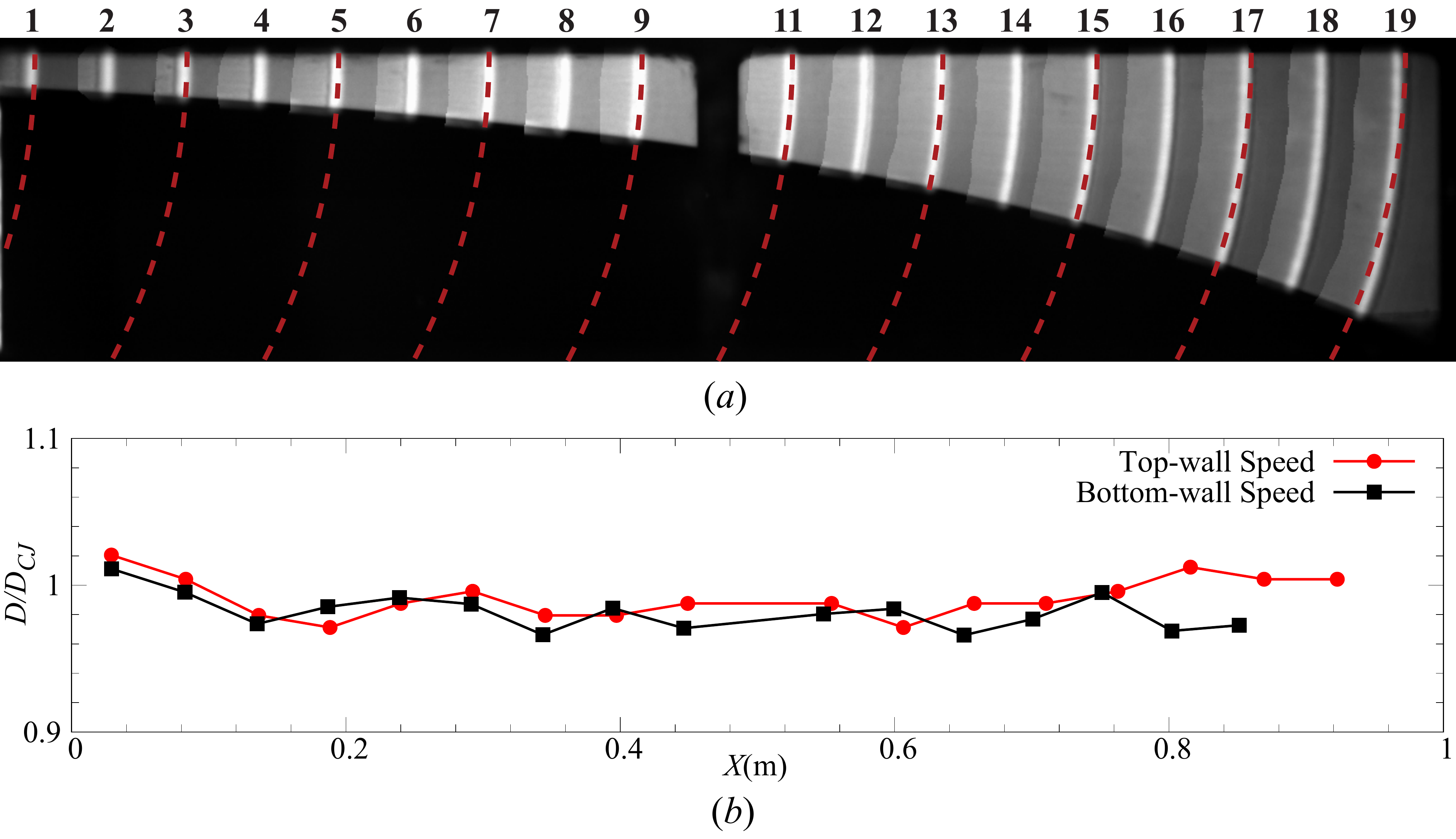}}
	\caption{Superposition of detonation fronts and the corresponding local speed profiles at different instants along the large ramp in the mixture of C${_2}$H${_4}$/3O$_2$ at an initial pressure of $p_0 = 8.3$ kPa. The red dashed lines in ($a$) denote arcs of circles with the expected curvature from the quasi-1D approximation. } \label{C2H4superimpose}  
\end{figure}

On the other hand, the theoretically expected arcs of circles of curvature predicted from the quasi-1D approximation \cite{xiao2019, radulescu2018} were obtained and compared with the real detonation fronts in experiments. These arcs of circles with the radius of $1/K = 0.46$ m are denoted by the red dashed lines in Fig.\ \ref{Superposition1}. The very good agreement between experiments and the theoretical expectations suggests the appropriateness of \textcolor{black}{the} quasi-1D assumption for detonations at the macro-scale in the specifically designed exponential geometry, as already demonstrated in detail by Xiao and Radulescu \cite{xiao2019} and Radulescu and Borzou \cite{radulescu2018}. The macro-scale quasi-steady assumption is further confirmed in Fig.\ \ref{C2H4superimpose}, illustrating the whole propagation process of  C${_2}$H${_4}$/3O$_2$ detonations along the large ramp at a higher initial pressure of $p_0 = 8.3$ kPa. Evidently, detonations propagated with a constant mean front curvature, indicated again by the excellent agreement with the red dashed arcs expected from the quasi-1D approximation. While for the minor deviations recognized near the end, they are due to the limitation of the quasi-1D approximation in designing the exponential ramp \cite{radulescu2018, xiao2019}. Moreover, the local speed profiles of both the top and bottom curved walls in Fig.\ \ref{C2H4superimpose}$b$ show very limited variations within 5\% of $D_{CJ}$, indicating detonations travelling at a constant speed. The measured global mean \textcolor{black}{propagation} speeds of the top and bottom walls differ within 1\% of $D_{CJ}$, \textcolor{black}{i.e., approximately $0.99D_{CJ}$ for the top wall and $0.98D_{CJ}$ for the bottom wall}. Therefore, it is reasonable for us to assume a quasi-steady detonation \textcolor{black}{at the macro-scale} with a constant mean lateral strain rate in the exponential channel.

\begin{figure}[]
	\centering
	{\includegraphics[width=0.95\textwidth]{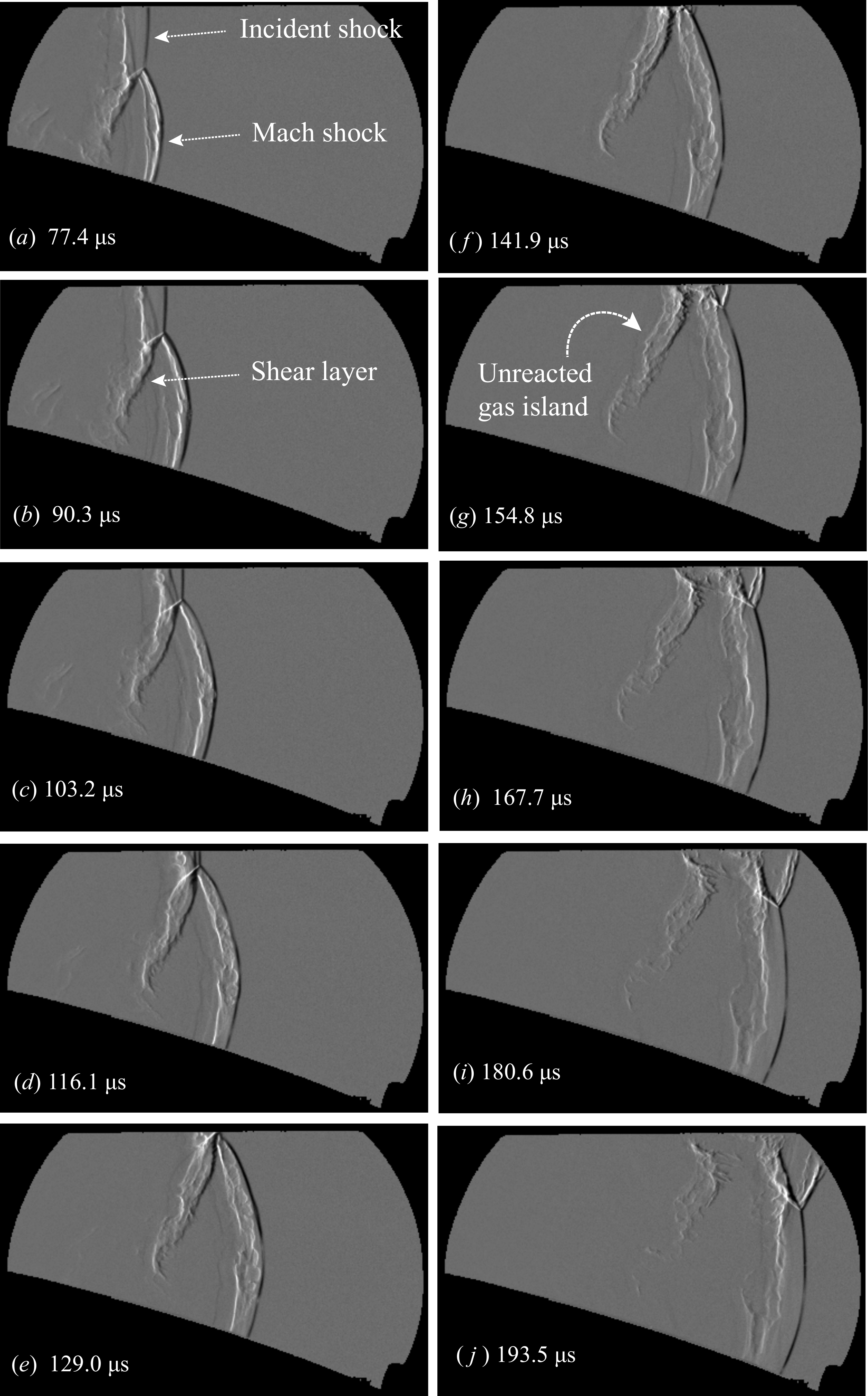}}
	\caption{Evolution of C${_2}$H${_4}$/3O$_2$ detonations along the large ramp at the initial pressure of 1.3 kPa.  } \label{C2H4-3O2-Evolution}  
\end{figure}

\begin{figure}[]
	\centering
	{\includegraphics[width=0.9\textwidth]{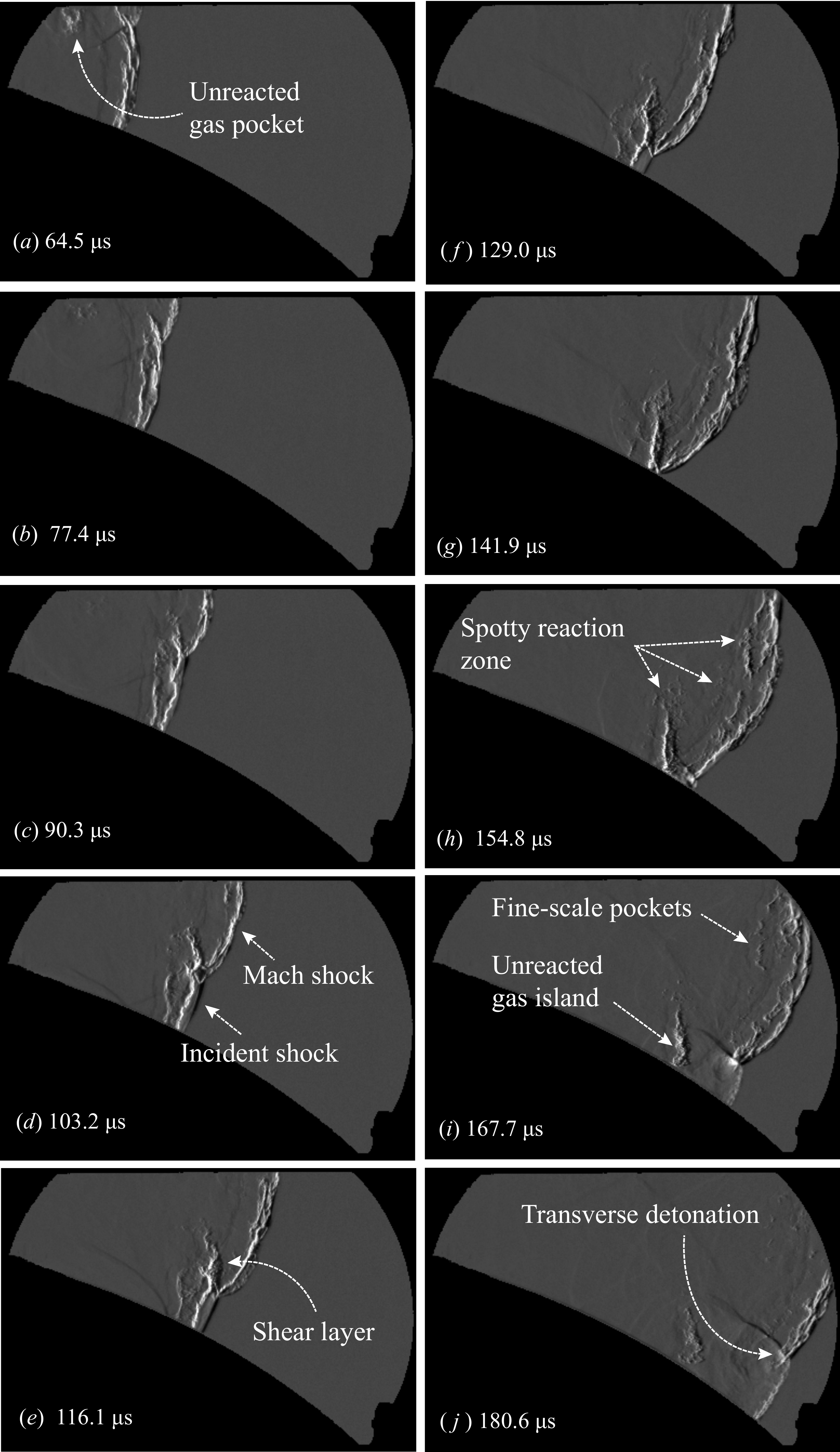}}
	\caption{Evolution of  C${_2}$H${_6}$/3.5O$_2$ detonations along the small ramp at the initial pressure of 4.6 kPa.  } \label{2C2H6-7O2-Evolution}  
\end{figure}

In addition, we also observed the propagation characteristics of near-limit detonations in mixtures of ethylene-oxygen (C${_2}$H${_4}$/3O$_2$) and ethane-oxygen (C${_2}$H${_6}$/3.5O$_2$), by reducing the kinetic sensitivity via lowering the initial pressures. In experiments of C${_2}$H${_4}$/3O$_2$, detonations can even self-sustain at very low initial pressures close to 1.0 kPa. Note that for these very low pressures, a driver gas of C${_2}$H${_4}$/3O$_2$ with a higher initial pressure was adopted in the initiation section, which was separated from the second part with a diaphragm shown in Fig.\ \ref{Experimental-setup}. Figures\ \ref{C2H4-3O2-Evolution} and \ref{2C2H6-7O2-Evolution}, respectively, illustrate the evolution of single-mode detonations in mixtures of C${_2}$H${_4}$/3O$_2$ and C${_2}$H${_6}$/3.5O$_2$. Compared to C${_2}$H${_4}$/3O$_2$ detonations, the C${_2}$H${_6}$/3.5O$_2$ detonation reaction zone structure appears to be more turbulent. The scenario of generating the tongue-shaped unreacted gas zone along the shear layer is qualitatively similar to that observed in methane-oxygen (CH${_4}$/2O$_2$) detonations in Fig.\ \ref{CH4-2O2-Evolution-2}. The significant unreacted gas islands were formed after the reflection of the singe triple point from the wall. These unreacted islands were again observed to be consumed more than four orders of magnitude faster than expected by the adiabatic shock-ignition \textcolor{black}{mechanism.} 

\begin{figure}[]
	\centering
	{\includegraphics[width=1.0\textwidth]{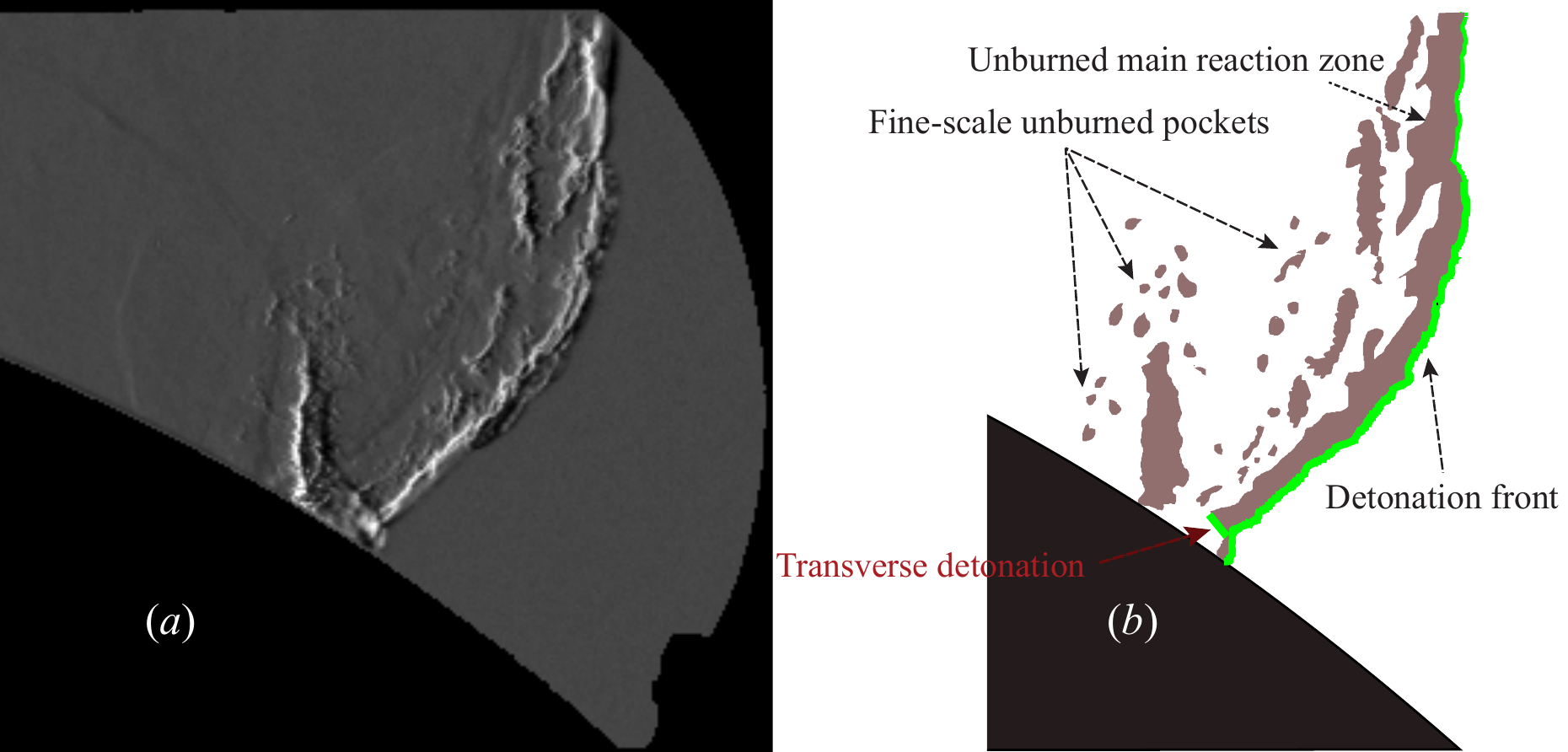}}
	\caption{Illustration of the scattered fine-scale unreacted pockets: ($a$) the spotty reaction zone structure of C${_2}$H${_6}$/3.5O$_2$ detonation in Fig.\ \ref{2C2H6-7O2-Evolution}$h$, and ($b$) sketch of the main features of the reaction zone.} \label{C2H6-pockets}  
\end{figure}

For C${_2}$H${_6}$/3.5O$_2$ detonations, after collision of the triple point with the bottom wall, a transverse detonation with intense luminosity was generated, as can be seen in Fig.\ \ref{2C2H6-7O2-Evolution}$i$ and Fig.\ \ref{2C2H6-7O2-Evolution}$j$, while absent in the C${_2}$H${_4}$/3O$_2$ detonation experiment. In Fig.\ \ref{2C2H6-7O2-Evolution}, we can also observe the presence of scattered fine-scale unburned pockets (with chemical activity) pinched off from the very spotty main \textcolor{black}{reaction} zone, as clearly shown in the sketch of Fig.\ \ref{C2H6-pockets}. These unreacted gas pockets were of different sizes and distributed randomly behind the main reaction zone. They were observed to be burned out in a very short time interval, approximately within two or three frames.

\subsection{The burning rate of unreacted gas pockets}
\label{The burning rate of unreacted gas pockets}

The above near-limit detonation experiments have clearly shown that a significant portion of fresh mixtures are engulfed in tongue-shaped unburned islands along the turbulent shear layers as well as in scattered fine-scale unreacted pockets, and that they react much faster than predicted by the shock-induced auto-ignition mechanism. As such, shock compression alone cannot explain the much faster combustion events observed in these unstable hydrocarbon-oxygen detonation experiments. On the other hand, quite \textcolor{black}{a few} studies, e.g., Refs. \cite{radulescu2005, radulescu2007, kiyanda2013, mahmoudi2015, maxwell2017, radulescu2018c}, have indicated that these pockets are consumed via surface turbulent flames. The present experiments, as illustrated in Figs.\ \ref{CH4-2O2-Evolution-1}, \ref{CH4-2O2-Evolution-2}, \ref{C2H4-3O2-Evolution}, and \ref{2C2H6-7O2-Evolution}, further \textcolor{black}{confirm} this burning mechanism through diffusive flames at very rough pocket boundaries, characteristic of significant hydrodynamic instabilities giving rise to enhanced turbulent mixing between the reacted and unreacted gases along the turbulent shear layers \cite{radulescu2005, radulescu2007,maxwell2017, radulescu2018c}. It is thus instructive to estimate the rate at which burning of these unburned fuel pockets occurs. Recently, Maxwell et al. \cite{maxwell2017} proposed an approximate method for evaluating this turbulent flame speed, by considering an isolated single pocket and measuring its characteristic size and the time interval required for being consumed. Specifically, such turbulent flame speed $S_t$ can be approximately calculated as
\begin{align}
S_t \approx \dfrac{L}{\Delta t}
\end{align}
where $\Delta t$ is the time interval, which the unreacted fuel pocket takes to be burned out.  $L$ is the characteristic length of the pocket, which can be estimated as
\begin{align}
L = \dfrac{V}{A_s}
\end{align}
where $V$ is the pocket's volume, which is obtained by tracing the pocket shape area, multiplied by the channel width in the third dimension.  $A_s$ is the peripheral surface area, estimated with the pocket perimeter multiplied also by the channel width in the third dimension.   

\begin{figure}[]
	\centering
	{\includegraphics[width=1.0\textwidth]{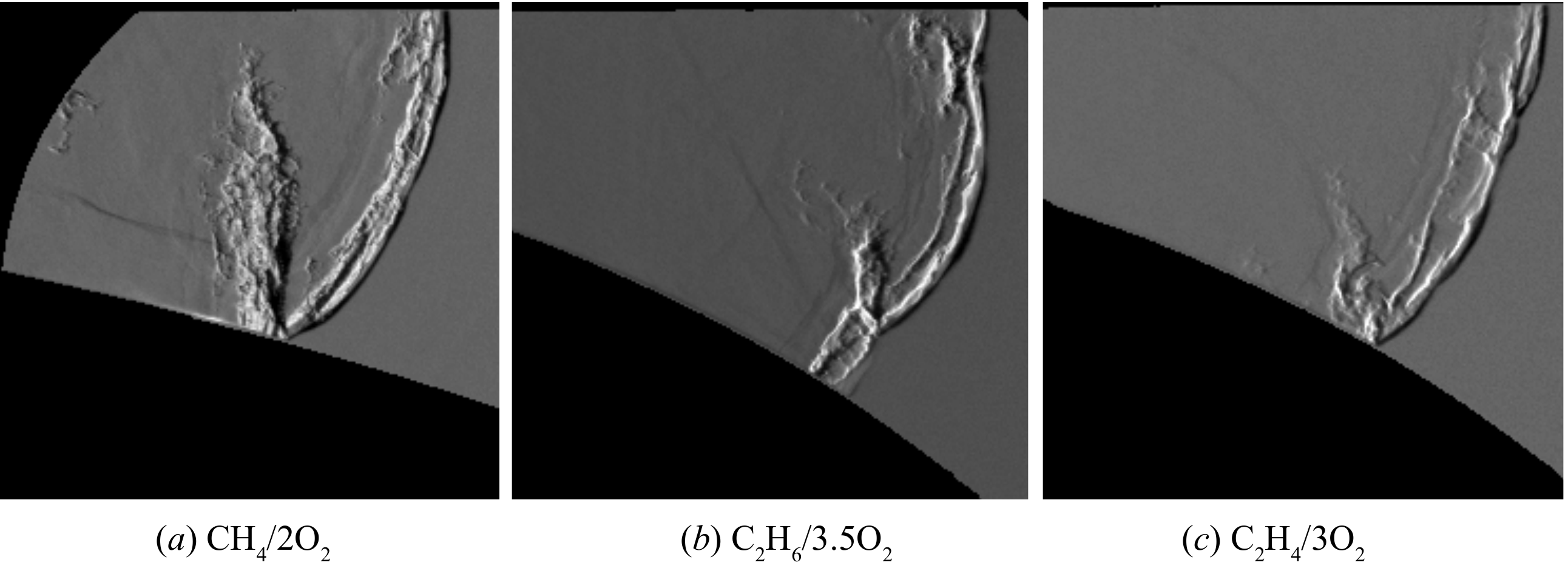}}
	\caption{Near-limit structures of ($a$) CH$_4$/2O$_2$ detonations along the large ramp in Fig.\ \ref{CH4-2O2-Evolution-2}e, ($b$) C$_2$H$_6$/3.5O$_2$ detonations along the small ramp at $p_0 = 4.5$ kPa, and ($c$) C$_2$H$_4$/3O$_2$ detonations along the small ramp at $p_0 = 1.8$ kPa.  } \label{CompareStructure}  
\end{figure}

Table \ref{St} shows the experimentally evaluated turbulent flame speeds $S_t$ for unburned fuel pockets starting in Fig.\ \ref{CH4-2O2-Evolution-2}$f$, Fig.\ \ref{C2H4-3O2-Evolution}$f$, and Fig.\ \ref{2C2H6-7O2-Evolution}$h$, respectively. These speeds vary approximately from 30 m/s to 70 m/s, which is smaller than that of \textcolor{black}{$110 \sim 120 $ m/s} obtained by Maxwell et al. \cite{maxwell2017} from the CH$_4$/2O$_2$ detonation experiment in a straight channel. Such discrepancy presumably originates from the much weaker incident shock ($D_{\textrm{avg}} \approx 0.5D_{CJ}$) in the present work resulting in lower post-shock temperatures than those of Maxwell et al. ($D_{\text{avg}} \approx 0.7 \sim 0.8D_{CJ}$) \cite{maxwell2017}, as well as in part from the stronger expansion waves behind the leading shock of detonations in the current diverging channel. Moreover, we also estimated the theoretical laminar flame burning velocity $S_L$ and CJ deflagration speed $S_{CJ}$ \cite{poludnenko2011, rakotoarison2019}, using the post-shock state of the incident shock that processed the unreacted gas pockets. The results in Table \ref{SL} show that, while the experimentally measured turbulent flame burning velocity $S_t$ is smaller than the CJ deflagration speed $S_{CJ}$ by a factor of two to three, it is two to seven times larger than the corresponding laminar flame burning velocity $S_L$. This agrees with the result of $S_t/S_L \approx 6.6 \sim 7.3$ evaluated by Maxwell et al. \cite{maxwell2017} in their CH$_4$/2O$_2$ detonation experiments. The decreasing sequence of $S_t/S_L$ from CH$_4$/2O$_2$ to C$_2$H$_6$/3.5O$_2$ and C$_2$H$_4$/3O$_2$ also signifies the decreasing level of turbulence effects. It is consistent with the qualitative comparison of the near-limit detonation reaction zone structures in these three mixtures, as shown in Fig.\ \ref{CompareStructure}, where CH$_4$/2O$_2$ detonation structures appear to be much rougher while the C$_2$H$_4$/3O$_2$ detonation is the opposite.

\begin{table}[]
	\centering
	\caption{The estimated turbulent flame burning velocity $S_t$ of the experimentally observed unburned fuel pockets in Figs. \ref{CH4-2O2-Evolution-2}, \ref{C2H4-3O2-Evolution}, and \ref{2C2H6-7O2-Evolution}.  }
	\begin{tabular}{ccccccc}
		\toprule
		Mixture & $p_0$ (kPa) & $V$ (mm$^3$) & $A_s$ (mm$^2$) & $L$ (mm) & $\Delta t$ ($\mu$s) & \multicolumn{1}{c}{$S_t$ (m/s)} \\
		\midrule
		CH$_4$/2O$_2$ & 12.7    & 39,000 & 5,900 & 6.6   & 103 & 64  \\
		C$_2$H$_6$/3.5O$_2$ & 4.6     & 11,000 & 2,900 & 3.8   & 78 & 49 \\
		C$_2$H$_4$/3O$_2$ & 1.2    & 31,000 & 6,000 & 5.2   & 155 & 34  \\
		\bottomrule
	\end{tabular}%
	\label{St}%
\end{table}%

\begin{table}[]
	\centering
	\caption{The laminar flame burning velocity $S_L$ and CJ deflagration speed $S_{CJ}$, evaluated from the post-shock state forming the unreacted gas pockets by considering the \textcolor{black}{detailed} chemistry \cite{Williams2014}. Note that $D_{\text{avg}}$ is the average speed of the weaker incident shock that \textcolor{black}{processed} the unreacted fuel pockets in in Figs. \ref{CH4-2O2-Evolution-2}, \ref{C2H4-3O2-Evolution}, and \ref{2C2H6-7O2-Evolution}.}
	\begin{tabular}{ccccccc}
		\toprule
		Mixture & $p_0$ (kPa) & $D_\text{{avg}}$/$D_{CJ}$  & $S_L$ (m/s)  & $S_{CJ}$ (m/s) & $ S_t / S_{CJ}$ & \multicolumn{1}{c}{$S_t / S_L$ } \\
		\midrule
		CH$_4$/2O$_2$ & 12.7  & 0.50       & 9.9  & 114.6  & 0.56 & 6.5 \\
		C$_2$H$_6$/3.5O$_2$ & 4.6   & 0.57      & 12.4  & 113.5 & 0.43  & 4.0  \\
		C$_2$H$_4$/3O$_2$ & 1.2   & 0.47     & 14.1  & 95.9   & 0.36 & 2.4 \\
		\bottomrule
	\end{tabular} 
	\label{SL}%
\end{table}%

\begin{figure}[]
	\centering
	{\includegraphics[width=1.0\textwidth]{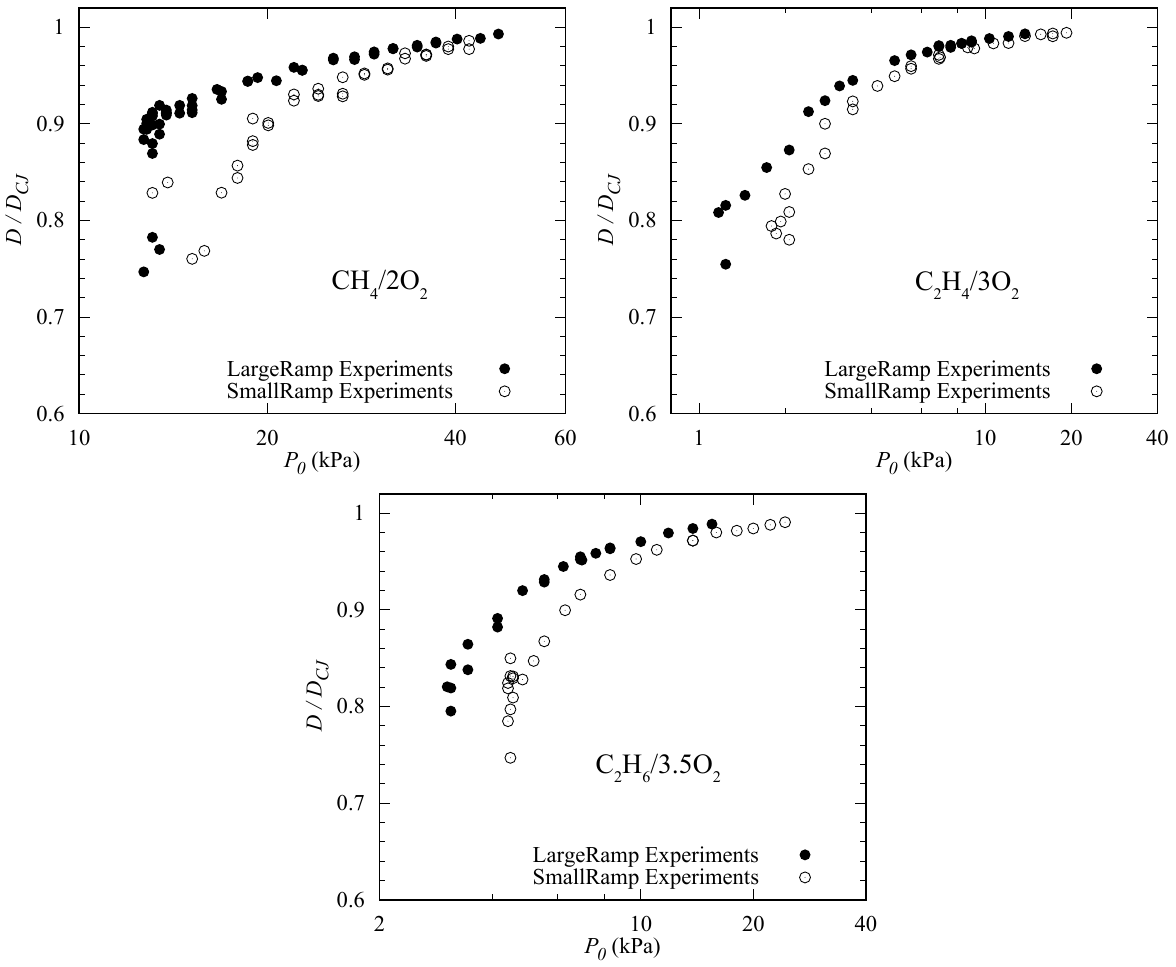}}
	\caption{The mean propagation speeds of detonations along the top wall of the whole ramp as a function of initial pressures in mixtures of CH${_4}$/2O$_2$, C${_2}$H${_4}$/3O$_2$, and C${_2}$H${_6}$/3.5O$_2$, respectively.} \label{DP}  
\end{figure}

\subsection{Velocity deficits}
Figure\ \ref{DP} summarizes the detonation mean propagation speeds measured along the top wall over the whole ramp, as a function of initial pressures. As a result of decreasing the mixture kinetic sensitivity through lowering the initial pressures, detonations exhibit larger velocity deficits. Since detonations along the large ramp experience a lower logarithmic area divergence rate than along the small ramp, they show a smaller velocity deficit as well as lower critical pressure, below which detonations fail to self-sustain. \textcolor{black}{ Of noteworthy is the relatively large spread of measured points recognized near the limit. This is presumably due to the highly stochastic property of these near-limit detonations, which gives rise to the measured average speeds of fluctuations.} Such near-limit stochastic phenomenon has been extensively reported in previous experiments, e.g., see Refs. \cite{Higgins-propane, lee2013, mevelhydrogen, xiao2019}.

\begin{figure}[]
	\centering
	{\includegraphics[width=1.0\textwidth]{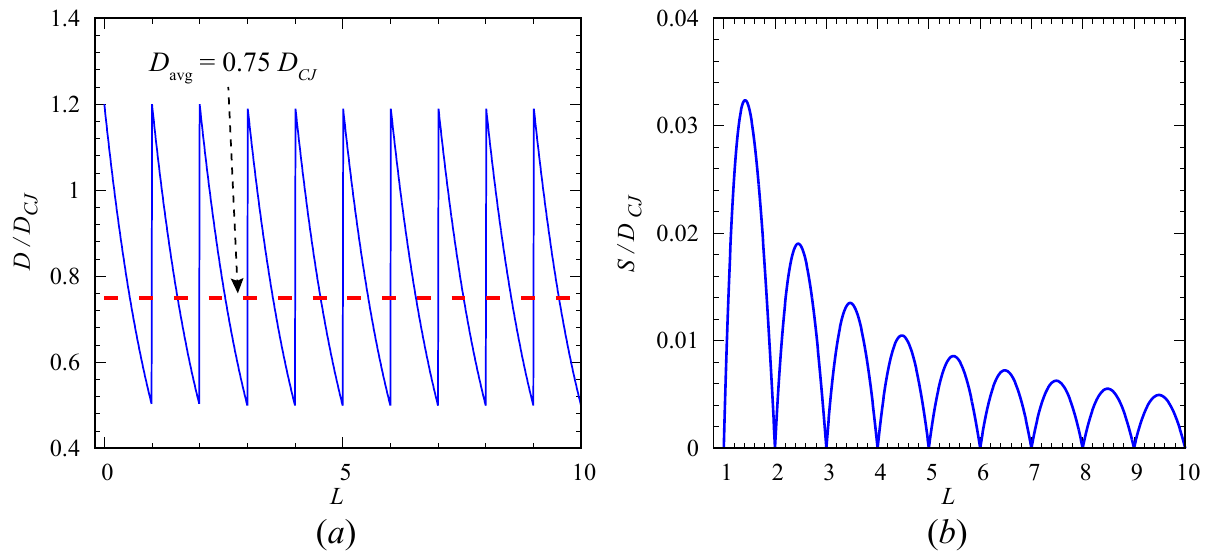}}
	\caption{\textcolor{black}{($a$) periodic detonation speed profiles  and ($b$) the standard deviation $S$ of the average speeds with respect to the physical channel length $L$. Note that $L$ was normalized by the characteristic cell length of detonations. }} \label{SpeedDeviation}  
\end{figure}

\textcolor{black}{We further analytically evaluated the uncertainty of the measured average speeds of the near-limit detonations with respect to the physical channel length, as shown in Fig.\ \ref{SpeedDeviation}. The abscissa denotes the length of channels, which was normalized by the detonation cell length. The periodic detonation speed evolution in each cell was depicted by an exponential function $D(x)/D_{CJ} = 1.2\text{exp}\left(-x/1.14\right)$ with $0 \le x \le 1$ , where $x$ is the non-dimensional distance. Such exponential relationship of detonation speeds in a cell can be justified by the linear relation between $D(x)$ and $\ode{D(x)}{t}$, as demonstrated by Jackson et al. \cite{jackson2019}. The resulting cyclic speed profiles are shown in Fig.\ \ref{SpeedDeviation}$a$, ranging from the maximum of 1.2 $D_{CJ}$ to the minimum of 0.5 $D_{CJ}$, characteristic of the amplitudes observed in the near-limit detonation experiments as shown in Fig.\ \ref{CH4-limit}$b$ and also detailed in Ref.\ \cite{xiao2019}. The average speed in a complete cycle is $D_\text{avg}$ = 0.75 $D_{CJ}$. Nevertheless, depending on the location, the measured average speeds over a specific channel can vary around 0.75 $D_{CJ}$. As such, we can obtain the standard deviation $S$ of these average speeds from $D_\text{avg}$ = 0.75 $D_{CJ}$ in terms of the channel length, as shown in Fig.\ \ref{SpeedDeviation}$b$. Clearly, as the channel length $L$ is longer than one cell, the standard deviation $S$ of the average speed over this channel is within 0.04 $D_{CJ}$. On the other hand, in the present work, the large ramp can host at least four complete cellular cycles for the near-limit detonations, as shown in Fig.\ \ref{CH4-limit}, while the small ramp can have two. Thus, the uncertainty of the experimentally measured average speeds of these near-limit detonations is within 0.02 $D_{CJ}$, which is negligible. In conclusion, the scatter reported in Fig.\ \ref{DP}, for example, for the near-limit detonations, is likely associated with the stochasticity of the detonation dynamics near the limit. Indeed, in diffraction experiments \cite{Higgins-propane,mevelhydrogen}, the limit itself was shown to be stochastic.}

\section{Discussion}
\label{Discussion}

\subsection{The experimental $D(\kappa)$ relationships}

\begin{figure}[]
	\centering
	{\includegraphics[width=1.0\textwidth]{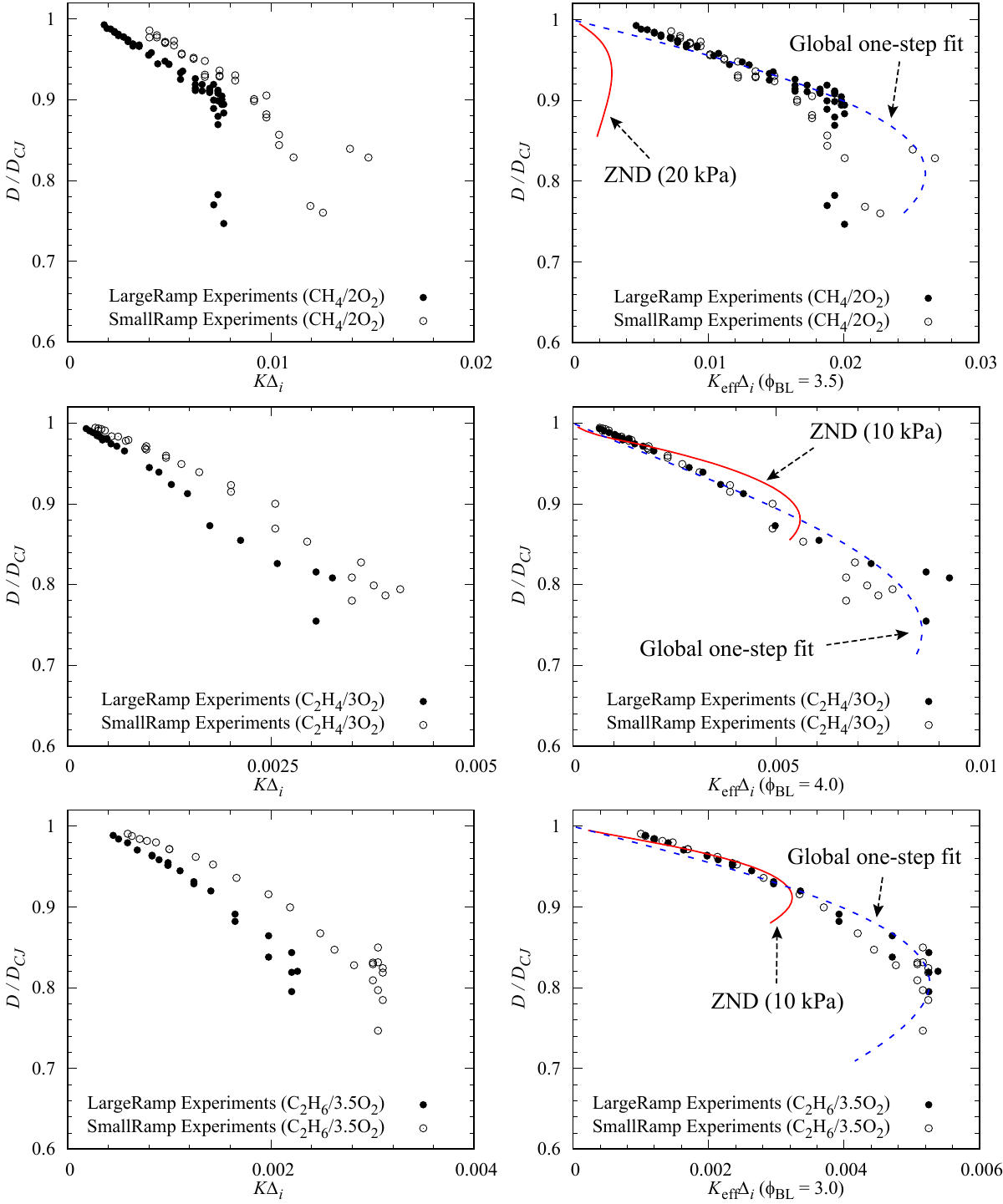}}
	\caption{The non-dimensional $D-K$ characteristic relationships obtained experimentally and predicted from the generalized ZND model with lateral strain rates \cite{klein1995, radulescu2018,xiao2019}. Note that the solid red line denotes the ZND model predictions with the \textcolor{black}{detailed reaction mechanism \cite{Williams2014}}, while the dashed blue line represents predictions using the empirical global one-step reaction model. All these ZND model predictions were calculated at the approximately medium pressure of the experimental range for each mixture.} \label{DKandDKeff}  
\end{figure}

The total flow divergence experienced by detonations propagating along ramps in this study includes two parts, the one due to the geometrical area divergence of the exponentially diverging channel and the other due to divergence of the flow rendered by the boundary layer growth on the side channel walls \cite{xiao2019,radulescu2018}.  The effective lateral flow divergence rate can thus be expressed as:
\begin{equation}
K_{\text{eff}} =\underbrace{\frac{1}{A}\ode{A}{x}}_{K}+ \mathrm{\phi_{BL}} \label{Keff}
\end{equation}
where $\mathrm{\phi_{BL}}$ represents the contribution of the boundary layer losses, and $K$ is the known logarithmic area divergence rate of the channel. One can refer to the work of Xiao and Radulescu \cite{xiao2019} for the theoretical justification of Eq.\ (\ref{Keff}). Instead of \textcolor{black}{modelling} the equivalent flow divergence rate $\mathrm{\phi_{BL}}$ of boundary layers, Radulescu and Borzou \cite{radulescu2018} proposed to evaluate this loss rate directly from experiments by analytically comparing the experimental data of two ramps, based on the following two assumptions: (1) detonations inside the exponential horn geometry of different divergence rates have the same constant $\mathrm{\phi_{BL}}$ since the channel's dimension  of the width is unchanged; (2) for the same mixture, it has a unique correlation between the velocity deficit and its loss \cite{klein1993relation, klein1995, he1994, yao1995}. As a result, the effective rate of total flow divergence $K_{\text{eff}}$ can be calibrated by collapsing together the experimental $D/D_{CJ}-K\Delta_i$ curves of both the large and small ramp experiments, by considering the boundary layer effects. As such, the loss rate $\mathrm{\phi_{BL}}$ due to boundary layers can be derived. Figure \ref{DKandDKeff} shows the experimentally obtained  $D-K$ curves characterizing the relationships between velocity deficits and losses for the mixtures involved in this study. Note that the abscissa is the non-dimensional loss obtained by multiplying the flow divergence rate with the \textcolor{black}{corresponding} CJ detonation induction zone length $\Delta_i$ (at the same initial pressure with the experiment), which was calculated by using Shepherd's SDToolbox \cite{lawson} with the San Diego chemical mechanism \cite{Williams2014}. In Fig.\ \ref{DKandDKeff}, the right column is the collapsed $D/D_{CJ}-K_{\text{eff}}\Delta_i$ correlation after calibrating the effective flow divergence rate $K_{\text{eff}}$ from the $D/D_{CJ}-K\Delta_i$ curves in the corresponding left graph by including the boundary layer effects. The \textcolor{black}{derived} constant boundary layer loss rates $\mathrm{\phi_{BL}}$ for CH${_4}$/2O${_2}$, C${_2}$H${_4}$/3O${_2}$, and C${_2}$H${_6}$/3.5O${_2}$ are 3.5 $\textnormal{m}^{-1}$, 4.0 $\textnormal{m}^{-1}$, and 3.0 $\textnormal{m}^{-1}$, respectively. \textcolor{black}{It is at present not clear what contributes to these different mean loss rates $\mathrm{\phi_{BL}}$ for different mixtures. Since the boundary-layer-induced losses depend on the growth of boundary layers inside the hydrodynamic reaction zone, $\mathrm{\phi_{BL}}$ is presumably controlled by the hydrodynamic thickness. More work is required for clarifying the relationship between the boundary layer losses and the hydrodynamic thickness.}

One interesting observation of the experimental $D/D_{CJ}-K_{\text{eff}}\Delta_i$ curves in Fig.\ \ref{DKandDKeff} is the extrapolation of the experimentally measured detonation speeds to zero divergence, the experimental data seem to extrapolate to speeds larger than the CJ values by a few percent. This result is the same with the earlier finding of Radulescu and Borzou \cite{radulescu2018} for the unstable C${_3}$H${_8}$/5O${_2}$ detonations. Recently, Damazo and Shepherd \cite{damazo2017} also reported the same phenomenon that, when extrapolated to zero losses, the experimental measurements were slightly larger than the CJ predictions by several percent. One can further refer to the review of Fickett and Davis \cite{fickett1979detonation} for more discussions and earlier references.

\textcolor{black}{While for comparisons of the experimentally obtained $D-K$ relationships with the generalized ZND model predictions using detailed chemical kinetics and the calibrated global one-step reaction model, as shown in Fig.\ \ref{DKandDKeff}, they will be discussed at length in the following parts of \ref{ZND-detailed} and \ref{ZND-onestep}, respectively.
}

\subsection{Comparisons with predictions of ZND detonations subject to lateral strain}
\label{ZND-detailed}
To start with, the experimental $D-K$ relationships were compared with those predicted from the generalized ZND model with lateral strain rates \cite{klein1995, xiao2019, radulescu2018}, as \textcolor{black}{denoted by the solid red lines} in Fig.\ \ref{DKandDKeff}. The generalized ZND model predictions of quasi-1D detonation dynamics with lateral flow divergence were calculated using the developed custom Python code \cite{radulescu2018} working under the framework of SDToolbox \cite{lawson} and Cantera \cite{goodwin2015}. The chemical kinetics was described by the San Diego reaction mechanism \cite{Williams2014}. These calculations were conducted at approximately the medium pressure of the experimental range for each mixture. Normalizing the varied divergence rates $K_{\text{eff}}$ by the CJ detonation induction zone length $\Delta_i$ at this medium pressure thus enables us to get the theoretical $D/D_{CJ}-K_{\text{eff}}\Delta_i$ curves. The comparisons between experiments and \textcolor{black}{detailed chemistry} predictions in Fig.\ \ref{DKandDKeff} show that while the generalized ZND model appears to predict relatively well the dynamics of  C${_2}$H${_4}$/3O$_2$ and C${_2}$H${_6}$/3.5O$_2$ detonations for small velocity deficits (approximately $D/D_{CJ} \ge 0.9$) and flow divergence, it considerably underpredicts the limiting flow divergence and the maximum velocity deficit. For CH${_4}$/2O$_2$ detonations, the prediction is much worse and completely underpredicts the experiments, where \textcolor{black}{the departure} can reach approximately one order of magnitude for the critical flow \textcolor{black}{divergence.} Evidently, these comparisons demonstrate that the unstable hydrocarbon-oxygen detonations in experiments can propagate with much larger lateral strain rates and velocity deficits than predicted by the steady ZND model, which concurs with the previous findings of Radulescu et al. \cite{radulescu2018, radulescu2002failure, radulescu2003propagation}.

Following the analysis of Short and Sharpe \cite{short2003}, Radulescu \cite{radulescu2003propagation} proposed the parameter $\chi$ for characterizing the detonation instability level, with detonations in mixtures of higher $\chi$ being more unstable to perturbations in reaction zones than those with lower $\chi$. The mathematical expression of  $\chi$ is 
\begin{equation}
\chi = \left(\dfrac{E_a}{RT_s}\right)\left(\dfrac{t_{ig}}{t_{re}} \right) \label{kai}
\end{equation} 
where $E_a/RT_s$ is the reduced activation energy, $T_s$ the post-shock temperature, $R$  the specific gas constant, and $t_{ig}/t_{re}$ the ratio of ignition to reaction time. Since the initial pressure does not change much the $\chi$ value of the CJ detonation in a specific mixture, we will adopt the initial pressures of the near-limit experiments for calculating the $\chi$ value, for characterizing the detonation instability level of each mixture in the subsequent sections.

\begin{figure}[htb!]
	\centering
	{\includegraphics[width=1.0\textwidth]{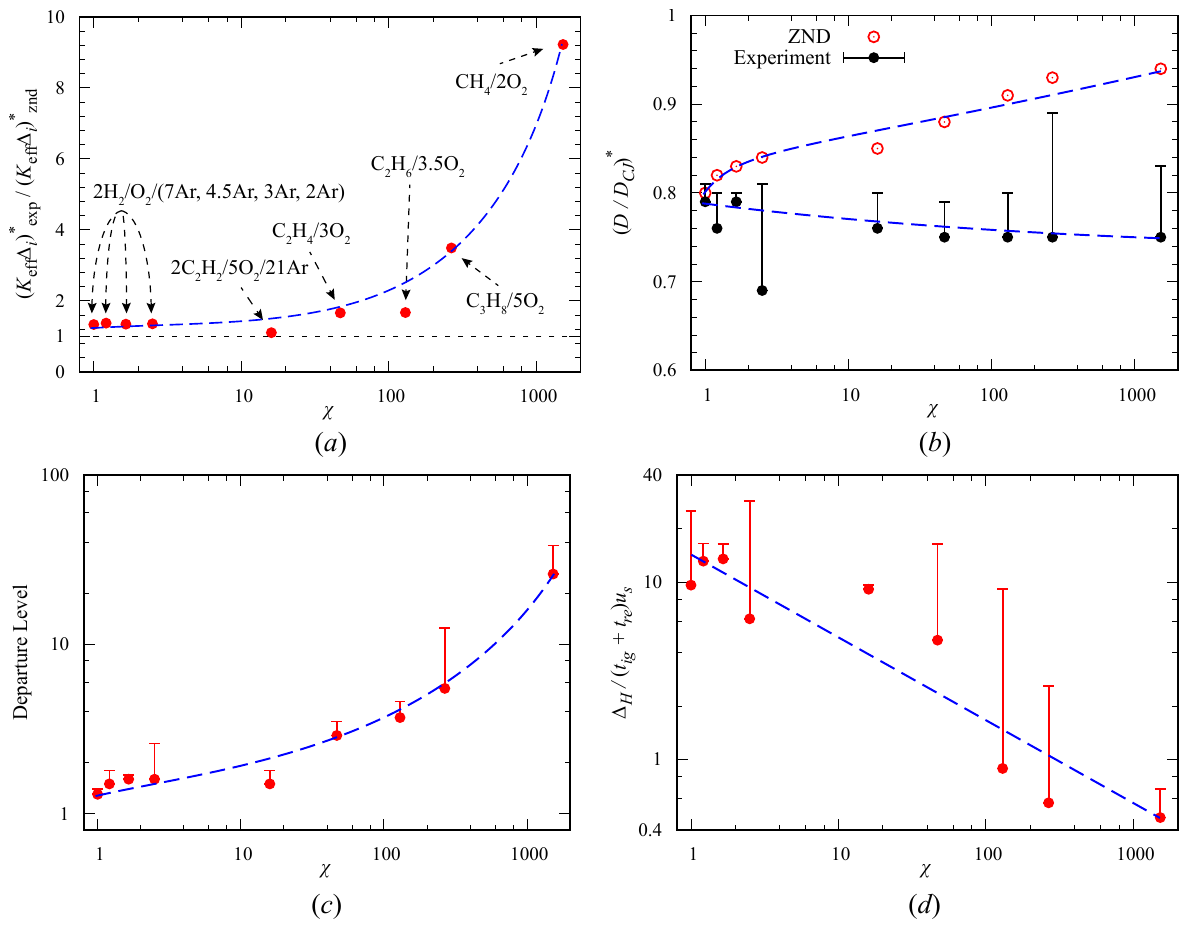}}
	\caption{Correlations of near-limit detonation parameters with the instability parameter $\chi$ for different mixtures: ($a$) ratio of the experimental limiting flow divergence to that predicted by the model, ($b$) the minimum mean propagation speed near the propagation limit, ($c$) the departure level (DL) defined by Eq.\ (\ref{DL}), and ($d$) ratio of the experimentally estimated reaction zone length $\Delta_H$ to the theoretical one. Note that relevant parameters for the mixtures of H$_2$/O$_2$/Ar, 2C${_2}$H${_2}$/5O${_2}$/21Ar, and C${_3}$H${_8}$/5O${_2}$ were adapted from the experiments reported by Radulescu et al. \cite{radulescu2018,xiao2019}. } \label{KaiCorrelation}  
\end{figure}

Figure\ \ref{KaiCorrelation} quantifies the departure of \textcolor{black}{experiments} from the theoretical predictions in terms of the near-limit detonation dynamics, i.e., the critical flow divergence $(K_{\text{eff}}\Delta_i)^*$ and the minimum mean propagation speed $(D/D_{CJ})^*$, for each mixture as a function of the instability level. Note that the relevant data of H$_2$/O$_2$/Ar, 2C${_2}$H${_2}$/5O${_2}$/21Ar, and C${_3}$H${_8}$/5O${_2}$ were adapted from the experiments already performed by our research group \cite{radulescu2018, xiao2019}. As the detonation instability $\chi$ value increases from the lowest of 2H${_2}$/O${_2}$/7Ar to the highest of CH${_4}$/2O$_2$, as illustrated in Fig.\ \ref{KaiCorrelation}$a$, departure of the experimentally obtained maximum lateral strain $(K_{\text{eff}}\Delta_i)_{\text{exp}}^*$ from the ZND model predicted counterpart $(K_{\text{eff}}\Delta_i)_{\text{znd}}^*$ shows a considerably increasing trend. While $(K_{\text{eff}}\Delta_i)_{\text{exp}}^*$/$(K_{\text{eff}}\Delta_i)_{\text{znd}}^*$ tends to 1 when $\chi$ approximates 0, it can reach one order of magnitude for CH${_4}$/2O$_2$ with $\chi \approx 1500$. Moreover, for the maximum velocity deficits in Fig.\ \ref{KaiCorrelation}$b$, the ZND model predicts the increasing $(D/D_{CJ})^*$ with the increase of $\chi$, due to the strong dependence of the critical velocity deficit on the activation energy \cite{he1994, klein1995, yao1995}. The higher activation energy theoretically permits a much smaller critical velocity deficit \cite{he1994, klein1995, yao1995}, and these reduced activation energies generally increase with $\chi$, as confirmed in Fig.\ \ref{G-Ea}$b$ and Fig.\ \ref{EaRTs}$a$, \textcolor{black}{which will be discussed later}. On the other hand, the experiments appear to show the opposite trend, with the measured minimum velocities being almost the same or even decreasing slightly with $\chi$. Noteworthy is that the error bar in Fig.\ \ref{KaiCorrelation}$b$ represents the minimum global average velocities obtained near the limit for both the large and small ramp experiments. Clearly, for the more stable 2H${_2}$/O${_2}$/7Ar detonation with $\chi$ close to 1, the generalized ZND model impressively predicts the same velocity deficit with the experiment, while for the highly unstable CH${_4}$/2O$_2$ detonation, the experimentally measured velocity deficit can be larger than the predicted deficit by \textcolor{black}{half an order of magnitude.}

Therefore, as the detonation instability level increases, \textcolor{black}{the degree of departure} of both the experimentally estimated critical lateral strain rates and maximum \textcolor{black}{velocity} deficits from the ZND model predicted values increases. Here we further propose the departure level (DL) defined by
\begin{align}
DL = \dfrac{(K_{\text{eff}}\Delta_i)_{\text{exp}}^*}{(K_{\text{eff}}\Delta_i)_{\text{znd}}^*}\times \dfrac{(1.0 - D/D_{CJ})_{\text{exp}}^*}{(1.0 - D/D_{CJ})_{\text{znd}}^*} \label{DL}
\end{align}
which denotes the departure of the critical flow divergence multiplied by that of the velocity deficit near the propagation limit. Figure\ \ref{KaiCorrelation}$c$ demonstrates such departure as a strong function of the instability level $\chi$. Thus, the instability $\chi$ appears to control the level of predictability of detonation dynamics in a specific mixture by the steady ZND model. 

In addition, we also evaluated the hydrodynamic reaction zone thickness $\Delta_H$ in the near-limit detonation experiments, by measuring the maximum length between the unreacted gas pockets (immediately before being consumed) and the leading shock. And we then compared it to the theoretical length $(t_{ig}+t_{re})u_s$, where the ignition time $t_{ig}$, reaction time $t_{re}$, and the post-shock particle velocity $u_s$ (in the shock-attached reference) were evaluated behind the shock of the experimentally measured average velocity, using the typical constant-volume explosion method. Compared to the less unstable system of 2H${_2}$/O${_2}$/7Ar, the highly unstable CH${_4}$/2O$_2$ detonation has a much shorter reaction zone length (observed from experiments) than the theoretical value, \textcolor{black}{as can be clearly seen from Fig.\ \ref{KaiCorrelation}$d$.} \textcolor{black}{It thus indicates} the existence of additional effects on significantly promoting the combustion in these unstable detonation experiments, \textcolor{black}{thereby} decreasing the characteristic reaction length.

\subsection{Empirical global one-step reaction models for the hydrodynamic \textcolor{black}{macroscopic} average description of cellular detonation dynamics }
\label{ZND-onestep}
A useful formulation of the detonation problem is seeking a \textcolor{black}{macroscopic} average description of the detonation structure, such that relevant empirical models can be developed for capturing the detonation dynamics at macro-scales. The major contribution of the present exponential ramp experiments relies on unambiguously providing a unique curve relating the detonation speed and the hydrodynamic streamline divergence, i.e., the characteristic $D-K$ relationship. Unfortunately, these relations in Fig.\ \ref{DKandDKeff} differ substantially from the generalized ZND model predictions obtained using \textcolor{black}{detailed} chemical kinetics for the mixture. These comparisons show that cellular detonations in experiments generally can propagate under global strain rates much larger than permissible for 1D waves. Indeed, this comes as no
surprise and clearly highlights the departure of the global energy release rate in the cellular detonation from that of the steady 1D ZND wave by a significant amount, as clearly indicated by the much shorter ignition time and length scales observed in experiments than the theoretical counterparts.

Despite the ZND model with \textcolor{black}{detailed} chemistry failing to correctly capture the response of detonations to the imposed flow divergence, in a practical manner, Radulescu and Borzou proposed using the exponential ramp experiments data to infer an empirical reaction model for the hydrodynamic average description of the macro-scale detonation dynamics \cite{radulescu2018}. Ironically, the hydrodynamic model is still the ZND model itself, with the fundamental difference that the global energy release rate represents the mean evolution extracted from experiments. The simplest description of the global energy release rate is the one-step reaction model \cite{radulescu2018}, \textcolor{black}{which follows the Arrhenius law as:} 
\begin{align}
\textcolor{black}{\omega_P = G\times k_A \left(1 - Y_P\right)\text{exp}\left(\dfrac{E_a}{RT}\right)} \label{one-step}
\end{align}
\textcolor{black}{where $\omega_P$ is the rate of production of product, $Y_P$ the mass fraction of the product, and $k_A$ the typical pre-exponential rate constant. $G$ is the scale factor, which is a tuning parameter. The initial pressure $p_0$, initial density $\rho_0$, and the ZND detonation induction time $t_{ig}$ were adopted as the normalization scales.} By doing the same exercise as Radulescu and Borzou \cite{radulescu2018}, the \textcolor{black}{scale factor $G$} and the effective activation energy \textcolor{black}{ $E_a/RT_0$} can be fitted such that the ZND model predicted $D-K$ curve matches with experiment.

Table\ \ref{Global-onstep-fit} lists all the thermo-chemical parameters required by the empirical reaction model for a global average description of detonation dynamics at \textcolor{black}{macro-scales}. $\gamma$ is the post-shock isentropic exponent of the CJ detonation, for correctly recovering the gas compressibility in the reaction zone. The heat release $Q$  was determined from the perfect gas relation \cite{fickett1979detonation} recovering the correct detonation Mach number. The real activation energy $E_a/RT_0$ (\textcolor{black}{Detailed}) was obtained from the slope of the \textcolor{black}{logarithmic} ignition delay with respect to inverse of the post-shock temperature with small perturbations. In these calculations, the \textcolor{black}{detailed kinetic mechanism} \cite{Williams2014} was applied. \textcolor{black}{Since the maximum velocity deficits of the near-limit detonations are a strong function of the activation energy  \cite{he1994, klein1995, yao1995},  the effective activation energy $E_a/RT_0$ (Empirical) can thus be readily fitted from these turning points, with their limiting lateral divergence rescaled by tuning the scale factor $G$ in Eq.\ (\ref{one-step}). } Of noteworthy is that the scale factor $G$, for the non-dimensional Arrhenius pre-exponential rate constant, actually represents the global reaction rate amplification factor. 
\begin{table}[]
	\centering
	\caption{Thermo-chemical parameters for the empirical one-step reaction model for a global description of cellular detonations, with the reaction rate amplification factor $G$ and the effective activation energy $E_a/RT_0$ fitted from the exponential ramp experiments in Fig.\ \ref{DKandDKeff}. }
	\begin{tabular}{cccccccc}
		\toprule
		Mixture & $p_0$ (kPa) & $\gamma $ & $\dfrac{Q}{RT_0}$ & $\dfrac{E_{a}}{RT_0}$(\textcolor{black}{Detailed}) & \textcolor{black}{$k_A$} & $\dfrac{E_{a}}{RT_0}$(Empirical)  & \multicolumn{1}{c}{$G$} \\
		\midrule
		CH$_4$/2O$_2$ & 20.0    & 1.17 & 64.5 &  70.0  & \textcolor{black}{15.0} & 24 & 8.5  \\
		C$_2$H$_4$/3O$_2$ & 10.0    & 1.17 & 73.1 &  30.4 &  \textcolor{black}{5.4} & 20 & 2.1  \\
		C$_2$H$_6$/3.5O$_2$ & 10.0     & 1.15 & 84.5 &  49.9 & \textcolor{black}{12.2} & 24 & 1.9 \\
		\bottomrule
	\end{tabular}%
	\label{Global-onstep-fit}%
\end{table}%

Since the resulting one-step reaction models can effectively describe the global rate of energy release in the cellular detonations, as clearly illustrated by the very good agreement between the experiments and global one-step predictions in Fig.\ \ref{DKandDKeff}, it is worth commenting on the magnitude of the two important fitting coefficients in Table\ \ref{Global-onstep-fit}. Obviously, the effective activation energy for a global description is much lower than the underlying chemical decomposition. Particularly for CH$_4$/2O$_2$ detonations, the reduction in energy can reach 66\%, signifying a dramatic change in the overall detonation dynamics \textcolor{black}{characteristics.} Alongside with these much lower effective activation energies, the global rates of energy release are greatly enhanced, as indicated by the reaction rate amplification factor $G$. It is thus clear that these unstable hydrocarbon-oxygen cellular detonations have a much more enhanced burning mechanism with the substantially suppressed thermal character of ignition, as compared to the laminar ZND wave.

\begin{figure}[]
	\centering
	{\includegraphics[width=1.0\textwidth]{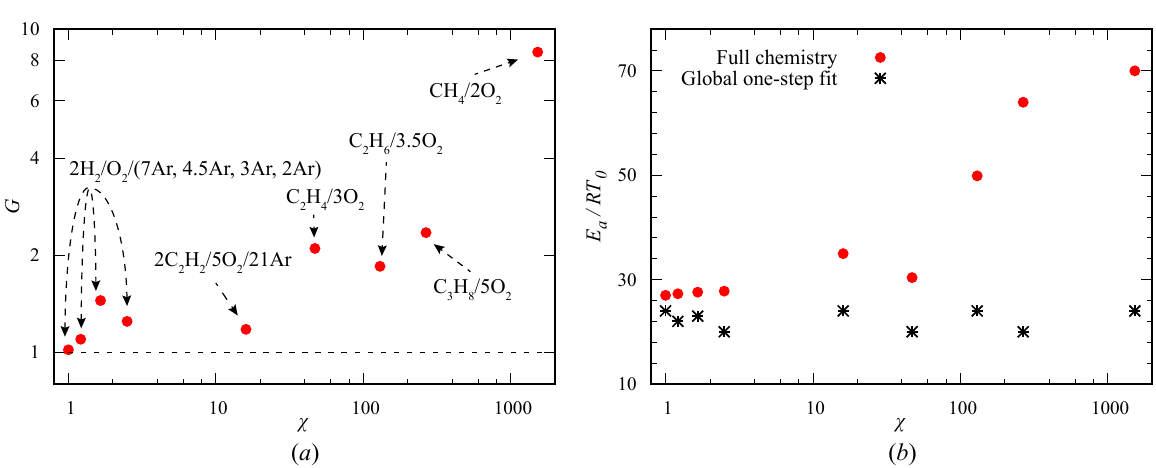}}
	\caption{The reaction rate amplification factor $G$ (scale factor for the non-dimensional pre-exponential rate constant) and the global activation energy ${E_a}/{RT_0}$, for the empirical one-step reaction model.    } \label{G-Ea}  
\end{figure}

In addition, we also performed the same exercise for the experiments of all the other mixtures done by our research group \cite{radulescu2018, xiao2019}. Figure \ref{G-Ea} shows the two important coefficients, i.e., the amplification factor $G$ and the effective activation energy, in terms of the instability parameter $\chi$ for various mixtures. Clearly, when $\chi$ tends to 0, $G$ becomes 1 and the effective activation energy is the same with the \textcolor{black}{activation energy obtained from detailed chemistry}. As a result, the macro-scale detonation dynamics of the more stable systems, such as 2H${_2}$/O${_2}$/7Ar, can be excellently captured by the ZND model, as detailed by Xiao and Radulescu \cite{xiao2019}. With the increase of $\chi$, both the reaction rate amplification effects and reduction of the real activation energy are much more intensified. The instability level $\chi$ appears to correlate well with the qualitative enhancement effects of the burning mechanism.

\textcolor{black}{While an effective method has been devised above for modelling the global mean reaction rate of unstable hydrocarbon-oxygen cellular detonations from experiments, this approach is not entirely satisfactory, since it misses the physical clarity. As such, the modeller has little insight into the cause and effect. Clearly, much more work is required for clarifying the physical relevancy of the present one-step fit for describing the global kinetics of detonation dynamics. Moreover, as the present work and previous ones \cite{radulescu2005, radulescu2007, kiyanda2013, mahmoudi2015, maxwell2017, radulescu2018c} have demonstrated the important roles of turbulent diffusion, future work should also be devoted to bottoms-up approaches \cite{radulescu2017icders} for rationalizing and identifying the most promising modelling avenue of the global-scale average exothermicity rate law for the hydrodynamic description of the detonation structure. }

\textcolor{black}{On the other hand, despite the limitation of the present one-step fit derived from experiments in terms of physical clarity, our motivation here is to provide simple rate laws from the well-defined experimental benchmark for effectively capturing the macroscopic detonation dynamics for macro-scale engineering applications. For example, one can possibly find a use of these laws in engineering-scale simulations of rotating detonation engines, where quasi-steady and stationary cellular detonation waves may appear and typically experience weak confinements. One can also refer to the report by Radulescu \cite{radulescu2017icders} for the usefulness of these laws in predicting the dynamics of detonations, such as initiation and failure. Clearly, much more efforts are required for further validations. }

\subsection{What contributes to enhancement of the global energy release rates in cellular detonations, as compared to a laminar ZND wave?}

The above analysis has quantitatively demonstrated that the highly unstable hydrocarbon-oxygen cellular detonations in experiments show a much more enhanced ignition mechanism, as compared to the laminar ZND wave. Thus, the significant question is, \it{what contributes to enhancement of the global reaction rates in cellular detonations?} \rm

Our clarification starts from the analysis of the relevant time and length scales. When assuming the ignition delay $t_{ig} \sim \textrm{exp}(E_a/RT_s)$, we can get
\begin{align}
\dfrac{t_{ig}}{t_{ig, CJ}} = \textrm{exp}\left\lbrace \dfrac{E_a}{RT_{s, CJ}}\left(\dfrac{T_{s, CJ}}{T_s} - 1\right) \right\rbrace \label{tig}
\end{align}
where $T_{s, CJ}$ is the post-shock temperature of the CJ detonation, while $T_s$ is that behind the detonation with velocity deficit. In the limit of strong shock assumption, i.e., $T_{s, CJ}/T_s \approx \left(D_{CJ}/D\right)^2$, and high activation energy, by performing a perturbation analysis of Eq.\ (\ref{tig}), we can then simplify it to
\begin{align}
\dfrac{t_{ig}}{t_{ig, CJ}} \approxeq \textrm{exp}\left\lbrace \left({2E_a}/{RT_{s, CJ}}\right)\left(1 - {D}/{D_{CJ}}\right) \right\rbrace \label{prediction}
\end{align}
which is the same with what we have recently obtained for cell sizes varying with respect to velocity deficits \cite{xiao2019effect}. It thus highlights the exponential sensitivity of the ignition time on velocity deficit, which is controlled by the global activation energy. 

\begin{figure}[]
	\centering
	{\includegraphics[width=1.0\textwidth]{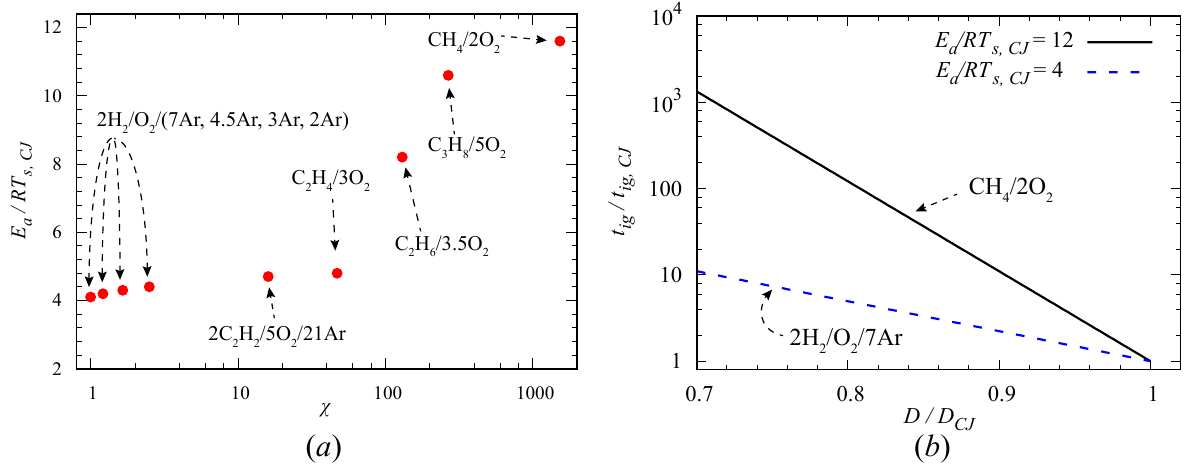}}
	\caption{Parameters of ($a$) the global activation energy $E_a/RT_{s, CJ}$ with respect to instability level $\chi$ for various mixtures, and ($b$) ignition time ratio predicted from Eq.\ (\ref{prediction}) for $E_a/RT_{s, CJ} = 12$ (CH$_4$/2O$_2$) and $E_a/RT_{s, CJ} = 4$ (2H$_2$/O$_2$/7Ar) as a function of velocity deficits.} \label{EaRTs}  
\end{figure}

As the highly unstable detonation has an evidently higher reduced activation energy, as shown in Fig.\ \ref{EaRTs}$a$, its response of ignition time to velocity deficit is much stronger. For example, at $D/D_{CJ}=0.7$, the ignition time of CH$_4$/2O$_2$ detonations is dramatically increased to approximately three orders of magnitude larger than the CJ detonation, as can be clearly seen from Fig.\ \ref{EaRTs}$b$. In cellular detonation experiments, e.g., in Fig.\ \ref{CH4-2O2-Evolution-2}, the weaker incident shock can even propagate at a local speed smaller than \textcolor{black}{0.5$D_{CJ}$.}  Due to the much longer ignition delays behind such weak portions of the leading shock, unreacted gases accumulate and are then pinched away from the front through expansion, giving rise to unreacted gas pockets distributed in the reaction zone, as clearly demonstrated in Fig.\ \ref{C2H6-pockets}. On the other hand, due to the relatively small activation energy, the weakly unstable 2H${_2}$/O${_2}$/7Ar detonation has a very limited variation of ignition time with respect to velocity deficits, as illustrated in Fig.\ \ref{EaRTs}$b$. As a result, very few significant pockets of unreacted gases have been observed in weakly unstable detonations.

Although these unreacted gas pockets are generated due to the much longer ignition delays behind weak incident shocks, their consumption time has been observed to be comparable to or smaller than the detonation cell timescale, suggesting their much faster reaction rates than the shock-induced auto-ignition mechanism. Moreover, visualization of the evolution of these pockets (see Figs.\ \ref{CH4-2O2-Evolution-2}, \ref{C2H4-3O2-Evolution}, and \ref{2C2H6-7O2-Evolution}) further shows that these pockets burn through the very rough pocket boundaries by turbulent surface flames, whose speed is approximately two to seven times larger than the corresponding laminar burning velocity. These much enhanced burning velocities highlight the important roles of turbulent mixing on the unreacted pocket surfaces and also between the reacted and unreacted gases along the turbulent shear layers \cite{radulescu2007,mahmoudi2015, maxwell2017}. Therefore, the much stronger sensitivity of ignition time on velocity deficits \textcolor{black}{associated with} the highly unstable cellular detonations \textcolor{black}{results in} significant unreacted pockets, whose combustion through diffusive surface flames is considerably promoted by turbulent mixing, thereby substantially suppressing the thermal ignition character and enhancing the global energy release rates.

\subsection{The $D/D_{CJ}-K_{\mathrm{eff}} \lambda$ curves}

\begin{figure}[htb!]
	\centering
	{\includegraphics[width=0.8\textwidth]{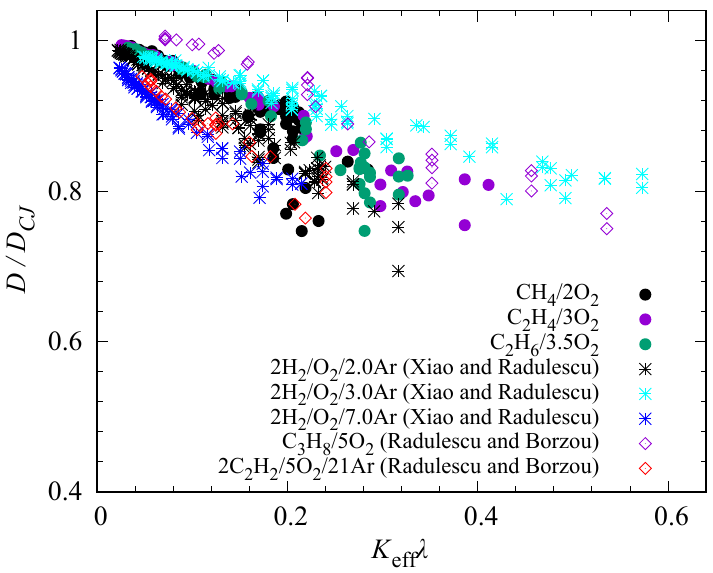}}
	\caption{The $D/D_{CJ}-K_{\text{eff}} \lambda$ relationships for various mixtures that have been done by the authors' group \cite{xiao2019, radulescu2018}.} \label{D-Klambda}  
\end{figure}

Finally, we also summarized the $D/D_{CJ}-K_{\text{eff}} \lambda$ relationships for various mixtures, as shown in Fig.\ \ref{D-Klambda}. These relations were obtained by using the already calibrated effective flow divergence rate $K_{\text{eff}}$ multiplied by the corresponding cell size $\lambda$. Note that $\lambda$ was obtained by appropriately fitting the cell size data of the corresponding mixture given in the Detonation Database of Caltech \cite{kaneshige1997}, \textcolor{black}{while the cell size of 2H$_2$/O$_2$/3Ar detonations (not available in database for the present experimental pressure range) was fitted from data obtained by Zhang et al. \cite{zhang2015}}. The fitting parameters of $\lambda$ are shown in Table \ref{cell-width}. \textcolor{black}{Of noteworthy is that the $D/D_{CJ}-K_{\text{eff}} \lambda$ curve of 2H$_2$/O$_2$/3Ar detonations goes beyond that of  2H$_2$/O$_2$/2Ar detonations in Fig.\ \ref{D-Klambda}, which disagrees with our previous finding that 2H$_2$/O$_2$/3Ar detonations fall between 2H$_2$/O$_2$/2Ar and 2H$_2$/O$_2$/7Ar detonations in terms of the $D/D_{CJ}-K_{\text{eff}} \Delta_i$ relationships \cite{xiao2019}. This inconsistency can be clarified due to the fitted cell size of 2H$_2$/O$_2$/3Ar detonations larger than that of 2H$_2$/O$_2$/2Ar by a factor of 2, as can be seen from the relations in Table \ref{cell-width}.} Taking into account the measurement uncertainty of cell sizes \cite{sharpe2011}, these $D/D_{CJ}-K_{\text{eff}} \lambda$ relations appear to follow a relatively good universality, as already found by Nakayama et al. \cite{nakayama2012, nakayama2013}. Such relatively good universality with $\lambda$ implicitly suggests that cell size is also a function of the detonation wave stability, since it is controlled by the reaction zone thickness of unstable detonations.

\begin{table}[htb!]
	\centering
	\caption{The fitting parameters for the cell size relations of  $\lambda = A*{p_0}^{-B}$ (mm).}
	\begin{tabular}{lccc}
		\toprule
		Mixture & $A$ & $B$ & \multicolumn{1}{c}{Reference}\\
		\midrule
		CH$_4$/2O$_2$ & 1231.2 & 1.37 & \cite{laberge1993propagation, abid1993oxidation, mcclenagan1988hydrazine, manzhalei1974measurement,knystautas1982critical}\\
		C$_2$H$_4$/3O$_2$ & 79.4 & 1.10 & \cite{abid1993oxidation, knystautas1982critical}\\
		C$_2$H$_6$/3.5O$_2$ & 261.2 & 1.28 & \cite{knystautas1982critical,bull1982detonation}\\
		C$_3$H$_8$/5O$_2$ & 190.8 & 1.15 & \cite{knystautas1982critical}\\
		2C$_2$H$_2$/5O$_2$/21Ar & 240.4 & 1.33 & \cite{strehlow1967a, desbordes1988, desbordes1993failure}\\
		2H$_2$/O$_2$/2Ar & 443.3 & 1.39 & \cite{strehlow1969transverse, barthel1974predicted}\\
		2H$_2$/O$_2$/3Ar & \textcolor{black}{999.1} & \textcolor{black}{1.38} & \textcolor{black}{\cite{zhang2015}}\\
		2H$_2$/O$_2$/7Ar & \textcolor{black}{1042.8} & \textcolor{black}{1.40} & \cite{strehlow1967a, barthel1974predicted}\\
		\bottomrule
		$p_0$ is the initial pressure in the dimensional unit of kPa.
	\end{tabular}
	\label{cell-width}
\end{table}

\section{Conclusion}
\label{Conclusion}
Experiments of detonations in three different hydrocarbon-oxygen mixtures with varied levels of cellular instability, i.e., CH$_4$/2O$_2$, C$_2$H$_4$/3O$_2$, and C$_2$H$_6$/3.5O$_2$, were conducted inside \textcolor{black}{exponentially} diverging channels. Visualization of the low-initial-pressure detonation reaction zone structures shows that these unstable detonations are characterized by the presence of significant unreacted gas pockets and turbulent spotty reaction zones.  The turbulent flame burning velocity of these pockets was evaluated, and found to be approximately 2 to 7 times larger than the corresponding laminar flame burning velocity, while smaller than the CJ deflagration speed by a factor of 2 to 3. Also, these experimentally visualized unreacted fuel pockets can be consumed more than four orders of magnitude faster than expected by the shock-induced ignition mechanism, thus confirming the previous observations highlighting the burning mechanism of these pockets through turbulent surface flames in unstable detonations \cite{radulescu2005, radulescu2007, kiyanda2013, maxwell2017}.

The characteristic $D-K$ relationships between the mean propagation speeds and dimensionless flow divergence were also obtained for each mixture, and compared with the generalized ZND model predictions of quasi-1D detonation dynamics in the presence of \textcolor{black}{lateral strain} using \textcolor{black}{detailed} chemical kinetics. Generally, these hydrocarbon-oxygen detonations in experiments were found to propagate under much larger limiting global divergence rates with higher maximum velocity deficits \textcolor{black}{than} predicted by the steady ZND model. It was also found that the detonation instability level $\chi$ appears to control the level of \textcolor{black}{departure} between experiments and the ZND model predictions in terms of the near-limit \textcolor{black}{detonation} dynamics.  The relatively good universality of the $D/D_{CJ}-K_{\text{eff}} \lambda$ relations implicitly suggests the dependence of cell size on the detonation wave stability.

By adopting the same method as Radulescu and Borzou \cite{radulescu2018}, the empirical global one-step reaction models were developed from \textcolor{black}{the data of experiments in exponentially diverging channels for} the hydrodynamic \textcolor{black}{macroscopic} average description of cellular detonation dynamics. The significance of the two important empirical coefficients, i.e., the reaction rate amplification factor $G$ and the effective activation energy, was \textcolor{black}{commented upon}. The considerably increased global reaction rates indicated by $G$ and much lower effective activation energies than the underlying \textcolor{black}{chemistry} decomposition highlight a much more enhanced burning mechanism of the highly unstable detonations through substantially suppressing the thermal character of the ignition, as compared to the laminar ZND wave.	The substantially enhanced global rates of energy release thus results in a much shorter hydrodynamic reaction zone length observed in experiments than the theory.

Finally, through analysis of the relevant time and length scales, the ignition time was found to follow the exponential dependence on the velocity deficit, which is controlled by the global activation energy. The much stronger sensitivity of the characteristic ignition time scales \textcolor{black}{on} velocity deficits \textcolor{black}{associated with} the highly unstable cellular detonations \textcolor{black}{results in} significant unreacted gas pockets. The burning of these pockets is further enhanced by turbulent mixing on flame surfaces and also between the reacted gases and unreacted gases along the turbulent shear layers, as already noted by Radulescu et al. \cite{radulescu2007} and Maxwell et al. \cite{maxwell2017}. As such, the overall thermal ignition character is suppressed and rates of global energy release are greatly enhanced. As a result, the laminar steady ZND model is not sufficient for capturing these unstable waves.
\section*{Acknowledgments}
\label{Acknowledgments}
The authors acknowledge the financial support from the Natural Sciences and Engineering Research Council of Canada (NSERC) through the Discovery Grant ``Predictability of detonation wave dynamics in gases: experiment and model development". \textcolor{black}{The authors would also like to thank Maxime La Fleche for help in conducting the experiments, Willstrong Rakotoarison for the script of calculating the CJ deflagration speeds, and Sebastien She-Ming Lau-Chapdelaine for the code of post-processing the experimental schlieren images.}

\bibliographystyle{elsarticle-num}
\bibliography{mybibfile} 

\begin{thebibliography}{10}
\expandafter\ifx\csname url\endcsname\relax
  \def\url#1{\texttt{#1}}\fi
\expandafter\ifx\csname urlprefix\endcsname\relax\def\urlprefix{URL }\fi
\expandafter\ifx\csname href\endcsname\relax
  \def\href#1#2{#2} \def\path#1{#1}\fi

\bibitem{xiao2019}
Q.~Xiao, M.~I. Radulescu, Dynamics of
  hydrogen\textendash{}oxygen\textendash{}argon cellular detonations with a
  constant mean lateral strain rate, Combustion and Flame 215 (2020) 437--457.

\bibitem{klein1995}
R.~Klein, J.~C. Krok, J.~E. Shepherd, Curved quasi-steady detonations:
  {{Asymptotic}} analysis and detailed chemical kinetics (May 1995).

\bibitem{radulescu2018}
M.~I. Radulescu, B.~Borzou, Dynamics of detonations with a constant mean flow
  divergence, Journal of Fluid Mechanics 845 (2018) 346--377.

\bibitem{fickett1979detonation}
W.~Fickett, W.~C. Davis, Detonation: {{Theory}} and {{Experiment}}, {Courier
  Corporation}, 2000.

\bibitem{lee_2008}
J.~H.~S. Lee, The Detonation Phenomenon, Cambridge University Press, 2008.

\bibitem{dabora1965}
E.~K. Dabora, J.~A. Nicholls, R.~B. Morrison, The influence of a compressible
  boundary on the propagation of gaseous detonations, Symposium (International)
  on Combustion 10~(1) (1965) 817--830.

\bibitem{reynaud2017a}
M.~Reynaud, F.~Virot, A.~Chinnayya, A computational study of the interaction of
  gaseous detonations with a compressible layer, Physics of Fluids 29~(5)
  (2017) 056101.

\bibitem{houim2017}
R.~W. Houim, R.~T. Fievisohn, The influence of acoustic impedance on gaseous
  layered detonations bounded by an inert gas, Combustion and Flame 179 (2017)
  185--198.

\bibitem{cho2017}
K.~Y. Cho, J.~R. Codoni, B.~A. Rankin, J.~Hoke, F.~Schauer, Effects of
  {{Lateral Relief}} of {{Detonation}} in a {{Thin Channel}}, in: 55th {{AIAA
  Aerospace Sciences Meeting}}, {{AIAA SciTech Forum}}, {American Institute of
  Aeronautics and Astronautics}, 2017.

\bibitem{mi2018b}
X.~C. Mi, A.~J. Higgins, C.~B. Kiyanda, H.~D. Ng, N.~Nikiforakis, Effect of
  spatial inhomogeneities on detonation propagation with yielding confinement,
  Shock Waves 28~(5) (2018) 993--1009.

\bibitem{chiquete2019}
C.~Chiquete, M.~Short, J.~J. Quirk, The {{Effect}} of {{Curvature}} and
  {{Confinement}} on {{Gas}}-{{Phase Detonation Cellular Stability}},
  Proceedings of the Combustion Institute 37~(3) (2019) 3565--3573.

\bibitem{nagura2013}
Y.~Nagura, J.~Kasahara, Y.~Sugiyama, A.~Matsuo, Comprehensive visualization of
  detonation-diffraction structures and sizes in unstable and stable mixtures,
  Proceedings of the Combustion Institute 34~(2) (2013) 1949--1956.

\bibitem{kudo2011oblique}
Y.~Kudo, Y.~Nagura, J.~Kasahara, Y.~Sasamoto, A.~Matsuo, Oblique detonation
  waves stabilized in rectangular-cross-section bent tubes, Proceedings of the
  Combustion Institute 33~(2) (2011) 2319--2326.

\bibitem{nakayama2012}
H.~Nakayama, T.~Moriya, J.~Kasahara, A.~Matsuo, Y.~Sasamoto, I.~Funaki, Stable
  detonation wave propagation in rectangular-cross-section curved channels,
  Combustion and Flame 159~(2) (2012) 859--869.

\bibitem{nakayama2013}
H.~Nakayama, J.~Kasahara, A.~Matsuo, I.~Funaki, Front shock behavior of stable
  curved detonation waves in rectangular-cross-section curved channels,
  Proceedings of the Combustion Institute 34~(2) (2013) 1939--1947.

\bibitem{rodriguez2019}
V.~Rodriguez, C.~Jourdain, P.~Vidal, R.~Zitoun, An experimental evidence of
  steadily-rotating overdriven detonation, Combustion and Flame 202 (2019)
  132--142.

\bibitem{short2019a}
M.~Short, C.~Chiquete, J.~J. Quirk, Propagation of a stable gaseous detonation
  in a circular arc configuration, Proceedings of the Combustion Institute
  37~(3) (2019) 3593--3600.

\bibitem{schwer2011numerical}
D.~Schwer, K.~Kailasanath, Numerical investigation of the physics of
  rotating-detonation-engines, Proceedings of the Combustion Institute 33~(2)
  (2011) 2195--2202.

\bibitem{lu2014}
F.~K. Lu, E.~M. Braun, Rotating {{Detonation Wave Propulsion}}: {{Experimental
  Challenges}}, {{Modeling}}, and {{Engine Concepts}}, Journal of Propulsion
  and Power 30~(5) (2014) 1125--1142.

\bibitem{paxson2018examination}
D.~E. Paxson, Examination of wave speed in rotating detonation engines using
  simplified computational fluid dynamics, in: 2018 AIAA Aerospace Sciences
  Meeting, 2018, p. 1883.

\bibitem{anand2019rotating}
V.~Anand, E.~Gutmark, Rotating detonation combustors and their similarities to
  rocket instabilities, Progress in Energy and Combustion Science 73 (2019)
  182--234.

\bibitem{zeldovich1950}
Y.~B. Zel'dovich, On the theory of the propagation of detonation in gaseous
  systems, {{NACA Technical Report}} 1261 (1950).

\bibitem{wood1954a}
W.~W. Wood, J.~G. Kirkwood, Diameter {{Effect}} in {{Condensed Explosives}}.
  {{The Relation}} between {{Velocity}} and {{Radius}} of {{Curvature}} of the
  {{Detonation Wave}}, The Journal of Chemical Physics 22~(11) (1954)
  1920--1924.

\bibitem{fay1959}
J.~A. Fay, Two-{{Dimensional Gaseous Detonations}}: {{Velocity Deficit}}, The
  Physics of Fluids 2~(3) (1959) 283--289.

\bibitem{tsuge1971}
S.~Tsug{\'e}, The {{Effect}} of {{Boundaries}} on the {{Velocity Deficit}} and
  the {{Limit}} of {{Gaseous Detonations}}, Combustion Science and Technology
  3~(4) (1971) 195--205.

\bibitem{dove1974}
J.~E. Dove, B.~J. Scroggie, H.~Semerjian, Velocity deficits and detonability
  limits of hydrogen-oxygen detonations, Acta Astronautica 1~(3) (1974)
  345--359.

\bibitem{murray1985}
S.~B. Murray, The {{Influence}} of {{Initial}} and {{Boundary Conditions}} on
  {{Gaseous Detonation Waves}}., Tech. Rep. DRES-SR-411, {Defence Research
  Establishment Suffield Ralston (Alberta)} (Sep. 1985).

\bibitem{gelfand1991}
B.~E. Gelfand, S.~M. Frolov, M.~A. Nettleton, Gaseous
  detonations\textemdash{{A}} selective review, Progress in Energy and
  Combustion Science 17~(4) (1991) 327--371.

\bibitem{agafonov1994}
G.~L. Agafonov, S.~M. Frolov, Computation of the detonation limits in gaseous
  hydrogen-containing mixtures, Combustion, Explosion and Shock Waves 30~(1)
  (1994) 91--100.

\bibitem{dionne2000}
J.-P. Dionne, H.~Dick~Ng, J.~H.~S. Lee, Transient development of
  friction-induced low-velocity detonations, Proceedings of the Combustion
  Institute 28~(1) (2000) 645--651.

\bibitem{ishii2002}
K.~Ishii, K.~Itoh, T.~Tsuboi, A study on velocity deficits of detonation waves
  in narrow gaps, Proceedings of the Combustion Institute 29~(2) (2002)
  2789--2794.

\bibitem{radulescu2002failure}
M.~I. Radulescu, J.~H. Lee, The failure mechanism of gaseous detonations:
  experiments in porous wall tubes, Combustion and Flame 131~(1-2) (2002)
  29--46.

\bibitem{radulescu2003propagation}
M.~I. Radulescu, The propagation and failure mechanism of gaseous detonations:
  experiments in porous-walled tubes, Ph.D. thesis, McGill University, Montreal
  (2003).

\bibitem{chao2009}
J.~Chao, H.~D. Ng, J.~H.~S. Lee, Detonability limits in thin annular channels,
  Proceedings of the Combustion Institute 32~(2) (2009) 2349--2354.

\bibitem{kitano2009}
S.~Kitano, M.~Fukao, A.~Susa, N.~Tsuboi, A.~K. Hayashi, M.~Koshi, Spinning
  detonation and velocity deficit in small diameter tubes, Proceedings of the
  Combustion Institute 32~(2) (2009) 2355--2362.

\bibitem{camargo2010}
A.~Camargo, H.~D. Ng, J.~Chao, J.~H.~S. Lee, Propagation of near-limit gaseous
  detonations in small diameter tubes, Shock Waves 20~(6) (2010) 499--508.

\bibitem{ishii2011a}
K.~Ishii, M.~Monwar, Detonation propagation with velocity deficits in narrow
  channels, Proceedings of the Combustion Institute 33~(2) (2011) 2359--2366.

\bibitem{zhang2015}
B.~Zhang, X.~Shen, L.~Pang, Y.~Gao, Detonation velocity deficits of
  {{H2}}/{{O2}}/{{Ar}} mixture in round tube and annular channels,
  International Journal of Hydrogen Energy 40~(43) (2015) 15078--15087.

\bibitem{gao2016}
Y.~Gao, B.~Zhang, H.~D. Ng, J.~H.~S. Lee, An experimental investigation of
  detonation limits in hydrogen\textendash{}oxygen\textendash{}argon mixtures,
  International Journal of Hydrogen Energy 41~(14) (2016) 6076--6083.

\bibitem{chinnayya2013}
A.~Chinnayya, A.~Hadjadj, D.~Ngomo, Computational study of detonation wave
  propagation in narrow channels, Physics of Fluids 25~(3) (2013) 036101.

\bibitem{sow2019}
A.~Sow, A.~Chinnayya, A.~Hadjadj, On the viscous boundary layer of weakly
  unstable detonations in narrow channels, Computers \& Fluids 179 (2019)
  449--458.

\bibitem{xiao2019effect}
Q.~Xiao, A.~Sow, B.~Maxwell, M.~I. Radulescu, Effect of boundary layer losses
  on 2d detonation cellular structures, arXiv preprint arXiv:1911.04617.

\bibitem{strehlow1967a}
R.~A. Strehlow, R.~Liaugminas, R.~H. Watson, J.~R. Eyman, Transverse wave
  structure in detonations, in: Symposium (International) on Combustion,
  Vol.~11, 1967, pp. 683--692.

\bibitem{monwar2007a}
M.~Monwar, Y.~Yamamoto, K.~Ishii, T.~Tsuboi, Detonation propagation in narrow
  gaps with various configurations, Journal of Thermal Science 16~(3) (2007)
  283--288.

\bibitem{mazaheri2015experimental}
K.~Mazaheri, Y.~Mahmoudi, M.~Sabzpooshani, M.~I. Radulescu, Experimental and
  numerical investigation of propagation mechanism of gaseous detonations in
  channels with porous walls, Combustion and Flame 162~(6) (2015) 2638--2659.

\bibitem{radulescu2017icders}
M.~I. Radulescu, The usefulness of a 1d hydrodynamic model for the detonation
  structure for predicting detonation dynamic parameters, in: 26th
  International Colloquium on the Dynamics of Explosions and Reactive Systems,
  Boston, USA, 2017.

\bibitem{zhang2013a}
B.~Zhang, C.~Bai, Critical energy of direct detonation initiation in gaseous
  fuel\textendash{}oxygen mixtures, Safety Science 53 (2013) 153--159.

\bibitem{radulescu2007}
M.~I. Radulescu, G.~J. Sharpe, C.~K. Law, J.~H.~S. Lee, The hydrodynamic
  structure of unstable cellular detonations, Journal of Fluid Mechanics 580
  (2007) 31--81.

\bibitem{maxwell2017}
B.~M. Maxwell, R.~R. Bhattacharjee, S.~S.~M. {Lau-Chapdelaine}, S.~a. E.~G.
  Falle, G.~J. Sharpe, M.~I. Radulescu, Influence of turbulent fluctuations on
  detonation propagation, Journal of Fluid Mechanics 818 (2017) 646--696.

\bibitem{radulescu2018c}
M.~I. Radulescu, A detonation paradox: {{Why}} inviscid detonation simulations
  predict the incorrect trend for the role of instability in gaseous cellular
  detonations?, Combustion and Flame 195 (2018) 151--162.

\bibitem{bhattacharjee2013}
R.~R. Bhattacharjee, Experimental {{Investigation}} of {{Detonation
  Re}}-initiation {{Mechanisms Following}} a {{Mach Reflection}} of a
  {{Quenched Detonation}}, M.{{Sc Thesis}}, University of Ottawa, Ottawa
  (2013).

\bibitem{dennis2014}
K.~Dennis, L.~Maley, Z.~Liang, M.~I. Radulescu, Implementation of large scale
  shadowgraphy in hydrogen explosion phenomena, International Journal of
  Hydrogen Energy 39~(21) (2014) 11346--11353.

\bibitem{radulescu2005}
M.~I. Radulescu, G.~J. Sharpe, J.~H.~S. Lee, C.~B. Kiyanda, A.~J. Higgins,
  R.~K. Hanson, The ignition mechanism in irregular structure gaseous
  detonations, Proceedings of the Combustion Institute 30~(2) (2005)
  1859--1867.

\bibitem{kiyanda2013}
C.~B. Kiyanda, A.~J. Higgins, Photographic investigation into the mechanism of
  combustion in irregular detonation waves, Shock Waves 23~(2) (2013) 115--130.

\bibitem{gamezo2000}
V.~N. Gamezo, A.~A. Vasil'ev, A.~M. Khokhlov, E.~S. Oran, Fine cellular
  structures produced by marginal detonations, Proceedings of the Combustion
  Institute 28~(1) (2000) 611--617.

\bibitem{Williams2014}
F.~A. Williams, Chemical {{Mechanism}}: {{Combustion Research Group}} at {{UC
  San Diego}},
  \url{https://web.eng.ucsd.edu/mae/groups/combustion/mechanism.html} (2014).

\bibitem{mahmoudi2015}
Y.~Mahmoudi, K.~Mazaheri, High resolution numerical simulation of triple point
  collision and origin of unburned gas pockets in turbulent detonations, Acta
  Astronautica 115 (2015) 40--51.

\bibitem{poludnenko2011}
A.~Y. Poludnenko, T.~A. Gardiner, E.~S. Oran, Spontaneous {{Transition}} of
  {{Turbulent Flames}} to {{Detonations}} in {{Unconfined Media}}, Physical
  Review Letters 107~(5) (2011) 054501.

\bibitem{rakotoarison2019}
W.~Rakotoarison, Y.~Vilende, M.~Radulescu, Model for chapman-jouguet
  deflagrations in open ended tubes with varying vent ratios, in: 27th
  International Colloquium on the Dynamics of Explosions and Reactive Systems,
  Beijing, China, 2019.

\bibitem{Higgins-propane}
J.~Loiseau, A.~J. Higgins, Statistical measurement of critical tube diameter,
  in: 21st International Colloquium on the Dynamics of Explosions and Reactive
  Systems, Poitiers, France, 2007.

\bibitem{lee2013}
J.~H.~S. Lee, A.~Jesuthasan, H.~D. Ng, Near limit behavior of the detonation
  velocity, Proceedings of the Combustion Institute 34~(2) (2013) 1957--1963.

\bibitem{mevelhydrogen}
R.~M{\'e}vel, Q.~Xiao, M.~Radulescu, Hydrogen-oxygen-argon detonation
  diffraction in a narrow channel, in: 26th International Colloquium on the
  Dynamics of Explosions and Reactive Systems, Boston, USA, 2017.

\bibitem{jackson2019}
S.~I. Jackson, C.~Chiquete, M.~Short, An intrinsic
  velocity\textendash{}curvature\textendash{}acceleration relationship for
  weakly unstable gaseous detonations, Proceedings of the Combustion Institute
  37~(3) (2019) 3601--3607.

\bibitem{klein1993relation}
R.~Klein, D.~S. Stewart, The relation between curvature, rate state-dependence,
  and detonation velocity, SIAM Journal on Applied Mathematics 53~(5) (1993)
  1401--1435.

\bibitem{he1994}
L.~He, P.~Clavin, On the direct initiation of gaseous detonations by an energy
  source, Journal of Fluid Mechanics 277 (1994) 227--248.

\bibitem{yao1995}
J.~Yao, D.~S. Stewart, On the normal detonation shock velocity-curvature
  relationship for materials with large activation energy, Combustion and Flame
  100~(4) (1995) 519--528.

\bibitem{lawson}
J.~Lawson, J.~Shepherd, Shock and detonation toolbox installation instructions,
  {{California Institute of Technology}}, Pasadena, CA (2019).

\bibitem{damazo2017}
J.~Damazo, J.~E. Shepherd, Observations on the normal reflection of gaseous
  detonations, Shock Waves 27~(5) (2017) 795--810.

\bibitem{goodwin2015}
D.~Goodwin, H.~Moffat, R.~Speth, Cantera: {{An Object}}-Oriented {{Software
  Toolkit}} for {{Chemical Kinetics}}, {{Thermodynamics}}, and {{Transport
  Processes}}. {{Version}} 2.2.0, 2015.

\bibitem{short2003}
M.~Short, G.~Sharpe, Pulsating instability of detonations with a two-step
  chain-branching reaction model: Theory and numerics, Combustion Theory and
  Modelling 7~(2) (2003) 401--416.

\bibitem{kaneshige1997}
M.~Kaneshige, J.~E. Shepherd, Detonation {{Database}},
  https://resolver.caltech.edu/CaltechGALCITFM:1997.008 (Sep. 1997).

\bibitem{sharpe2011}
G.~J. Sharpe, M.~I. Radulescu, Statistical analysis of cellular detonation
  dynamics from numerical simulations: One-step chemistry, Combustion Theory
  and Modelling 15~(5) (2011) 691--723.

\bibitem{laberge1993propagation}
S.~Laberge, R.~Knystautas, J.~Lee, Propagation and extinction of detonation
  waves in tube bundles, Progress in Astronautics and Aeronautics 153 (1993)
  381--396.

\bibitem{abid1993oxidation}
S.~Abid, G.~Dupre, C.~Paillard, Oxidation of gaseous unsymmetrical
  dimethylhydrazine at high temperatures and detonation of udmh/o$_2$ mixtures,
  Progress in Astronautics and Aeronautics 153 (1993) 162--181.

\bibitem{mcclenagan1988hydrazine}
M.~Pedley, C.~Bishop, F.~Benz, C.~Bennett, R.~McClenagan, D.~Fenton,
  R.~Knystautas, J.~Lee, O.~Peraldi, G.~Dupre, J.~Shepherd, Hydrazine vapor
  detonations, Progress in Astronautics and Aeronautics 114 (1988) 45--63.

\bibitem{manzhalei1974measurement}
V.~Manzhalei, V.~Mitrofanov, V.~Subbotin, Measurement of inhomogeneities of a
  detonation front in gas mixtures at elevated pressures, Combustion,
  Explosion, and Shock Waves 10~(1) (1974) 89--95.

\bibitem{knystautas1982critical}
R.~Knystautas, J.~Lee, C.~Guirao, The critical tube diameter for detonation
  failure in hydrocarbon-air mixtures, Combustion and Flame 48 (1982) 63--83.

\bibitem{bull1982detonation}
D.~Bull, J.~Elsworth, P.~Shuff, E.~Metcalfe, Detonation cell structures in
  fuel/air mixtures, Combustion and Flame 45 (1982) 7--22.

\bibitem{desbordes1988}
D.~Desbordes, Transmission of overdriven plane detonations: critical diameter
  as a function of cell regularity and size, Progress in Astronautics and
  Aeronautics 114 (1988) 170--185.

\bibitem{desbordes1993failure}
D.~Desbordes, C.~Guerraud, L.~Hamada, H.~Presles, Failure of the classical
  dynamic parameters relationships in highly regular cellular detonation
  systems, Progress in Astronautics and Aeronautics 153 (1993) 347--359.

\bibitem{strehlow1969transverse}
R.~Strehlow, C.~D. Engel, Transverse waves in detonations: I{I}-structure and
  spacing in {H}$_2$-{O}$_2$, {C}$_2${H}$_4$-{O}$_2$, and {C}{H}$_4$-{O}$_2$
  systems., AIAA journal 7~(3) (1969) 492--496.

\bibitem{barthel1974predicted}
H.~O. Barthel, Predicted spacings in hydrogen-oxygen-argon detonations, The
  Physics of Fluids 17~(8) (1974) 1547--1553.

\end{thebibliography}
\end{document}